\documentclass[useAMS,usenatbib,fleqn,twocolumn]{mnras}
\usepackage{psfrag}
\usepackage{subfig}
\usepackage{graphicx}	
\usepackage{amsmath}	
\usepackage{amssymb}	
\usepackage{multicol}        
\usepackage{bm}		
\usepackage{hyperref}
\usepackage{xcolor}
\usepackage{float} 
\usepackage{ulem}
\usepackage{lastpage}

\usepackage[T1]{fontenc}
\usepackage{ae,aecompl}

\usepackage{breqn}

\title[Bispectrum multipoles with massive neutrinos]{Redshift-space galaxy bispectrum in presence of massive neutrinos: \\
A multipole expansion approach for \textit{Euclid}}

\author[S. Pal et al.]{Sourav Pal,$^{1}$\thanks{E-mail:soupal1729@gmail.com}
Debanjan Sarkar,$^{2,3}$\thanks{E-mail:debanjan.sarkar@mcgill.ca}
Rickmoy Samanta,$^{4}$\thanks{E-mail:rickmoysamanta@gmail.com}
Supratik Pal $^{1}$\thanks{E-mail:supratik@isical.ac.in}
\\\\
$^{1}$ Physics and Applied Mathematics Unit, Indian Statistical Institute, 203 B.T. Road, Kolkata 700108, India\\
$^{2}$ Department of Physics and Trottier Space Institute, McGill University, QC H3A 2T8, Canada\\
$^{3}$ Ciela—Montreal Institute for Astrophysical Data Analysis and Machine Learning, QC H2V0B3, Canada\\
$^{4}$ Department of Physics, Birla Institute of Technology and Science Pilani, Hyderabad 500078, India
}

\begin{document}
\label{firstpage}
\pagerange{\pageref{firstpage}--\pageref{lastpage}}
\maketitle

\begin{abstract}
Massive neutrinos imprint distinctive signatures on the evolution of cosmic structures, notably suppressing small-scale clustering. We investigate the impact of massive neutrinos on the galaxy bispectrum in redshift-space, adopting a spherical harmonic multipole decomposition $B_L^m(k_1, \mu, t)$,  that captures the full angular dependence. We develop an analytical and numerical framework incorporating neutrino-corrected perturbation theory kernels and redshift-space distortions. Our results demonstrate that the linear triangle configurations are particularly sensitive to massive neutrinos, with deviations reaching up to $\sim 2\%$ for a total mass $\sum m_\nu = 0.12\,\mathrm{eV}$. To assess detection prospects in galaxy surveys like \textit{Euclid}, we compute the signal-to-noise ratio (SNR) for individual multipoles, including the effects of Finger-of-God damping and shot noise.  The neutrino-induced signatures in $B_0^0$ and $B_2^0$ are found to be detectable with SNR $\gtrsim 5$ across a range of configurations, even after accounting for small-scale suppression. Higher-order multipoles such as $B_2^1$ and $B_2^2$ are moderately sensitive, with SNR $\gtrsim$ ($2-3$) in squeezed limits, while hexadecapole moments  are more suppressed but still exhibit measurable signals at high $k_1$. Additionally, the SNR generally increases with wave number $k_1$, particularly for squeezed and stretched triangles, suggesting that access to smaller scales significantly enhances detection prospects. Our study highlights the potential of the redshift-space bispectrum multipoles as sensitive probes of massive neutrinos, complementing traditional power spectrum analyses, and underscores the importance of angular information and higher-order statistics for galaxy surveys.
\end{abstract}

\begin{keywords}
cosmology: theory -- large-scale structure of Universe --  methods: statistical
\end{keywords}



\section{Introduction}
\label{sec:introduction}

Large-scale galaxy redshift surveys serve as a powerful tool for probing the growth of structures and hence the underlying cosmological model. In galaxy redshift surveys, the observed positions of galaxies are affected not only by the Hubble expansion but also by their peculiar velocities along the line-of-sight. This effect, known as redshift-space distortion (RSD), introduces anisotropies in the observed clustering statistics of galaxies. On large scales, due to the well-known Kaiser effect \citep{Kaiser:1987qv}, coherent infall  of galaxies into over-dense regions leads to an enhancement of the clustering amplitude along the line-of-sight.
On the other hand, on small scales, random motions in virialized structures suppress clustering, giving rise to the so-called Finger of God (FoG) effect \citep{Jackson:1971sky}. Together, these phenomena imprint distinct angular dependence on both the galaxy power spectrum and bispectrum in redshift-space.

The mutipoles of the redshift-space power spectrum \citep{Hamilton:1997zq} have been widely studied in the literature to constrain (i) the logarithmic growth rate $f$ \citep{Peacock:2001gs,Hawkins:2002sg,Guzzo:2008ac,2011ApJ...726....5O,Okumura:2023pxv}, (ii) neutrino mass \citep{Upadhye:2017hdl,DESI:2025ejh,Verdiani:2025znc} as well as (iii) dark energy and modified gravity theories \citep{Linder:2007nu,delaTorre:2016rxm,Rodriguez-Meza:2023rga,Valogiannis:2019nfz}. 
While the power spectrum is not capable of capturing the non-Gaussianities arising from gravitational clustering, bispectrum — the simplest higher order statistic beyond the power spectrum  — encodes additional information about non-linear gravitational evolution and galaxy bias. 
This further makes it a powerful tool to break degeneracies between cosmological parameters \citep{Scoccimarro:1999ed,Verde:1998zr,Gil-Marin:2014sta,Gil-Marin:2016wya,Nandi:2024cib}. The bispectrum is also affected by the RSD, which necessitates proper modelling in the redshift-space to accurately quantify the induced anisotropy. 
\citet{Scoccimarro:1999ed} made an early attempt to characterize RSD-induced anisotropy using spherical harmonics by focusing on the monopole and a single quadrupole component ($L = 2, m = 0$), an approach that was later followed by \citet{Hashimoto:2017klo}. \citet{Nan:2017oaq}  extended this by deriving approximate expressions for higher multipoles using the halo model, albeit for a limited set of triangle configurations. 
Subsequent works showed that adding bispectrum monopole information to the power spectrum significantly improves constraints on cosmological parameters \citep{Yankelevich:2018uaz,Philcox:2022frc,Philcox:2022lbx}.

Recent developments in computational techniques, such as Fast Fourier Transform (FFT)-based  bispectrum estimators \citep{Sugiyama:2018yzo,Shaw:2021pgy,Benabou:2023ldb} and modal decomposition approaches \citep{Byun:2022rvn}, have enabled efficient extraction of this anisotropic information. However, spherical harmonic basis still remains as the simplest and natural one to adopt. In this work, we extend the spherical harmonic decomposition framework introduced in \citet{Bharadwaj:2020wkc,Mazumdar:2020bkm,Mazumdar:2022ynd} to account for the impact of massive neutrinos, allowing for a detailed investigation of their effect on redshift-space bispectrum anisotropies.

The motivation for including massive neutrinos arises from both particle physics and cosmology. Neutrino oscillation experiments have conclusively established that neutrinos have mass, pointing to physics beyond the Standard Model \citep{IceCubeCollaboration:2023wtb,SNO:2002tuh,NOvA:2023uxq}. However, these experiments only provide lower bounds on the total neutrino mass through the measurement of mass squared differences  but cannot resolve the mass hierarchy. In the normal hierarchy (NH), the lightest states are nearly degenerate, requiring $\sum m_\nu \gtrsim 0.06 \, {\rm eV}$, while the inverted hierarchy (IH) demands $\sum m_\nu \gtrsim 0.10 \, {\rm eV}$ \citep{SajjadAthar:2021prg,Capozzi:2021fjo,Hannestad:2016fog}. Cosmology offers a complementary probe of neutrino mass through their subtle but cumulative effects on the expansion history and structure formation \citep{Dolgov:2002wy, Lesgourgues:2012uu}. The current tightest upper bound from cosmological observations is $\sum m_\nu <0.072\, {\rm eV}$, derived from Planck cosmic microwave background (CMB) data in combination with baryon acoustic oscillation (BAO) measurements from DESI \citep{ DESI:2024mwx}. This would effectively rule out the inverted mass hierarchy and raise concerns regarding the consistency of the standard hierarchy. 
Recent studies have also explored the possibility of negative ‘effective’ neutrino mass as a phenomenological solution to this issue \citep{Green:2024xbb, Craig:2024tky, Elbers:2024sha}. However, these results are not conclusive, since they primarily stem from lensing anomaly as pointed out in \citep{Naredo-Tuero:2024sgf,Allali:2024aiv,RoyChoudhury:2024wri,RoyChoudhury:2025dhe}.
To maintain robustness in our analysis, we therefore adopt the more conservative bound of $\sum m_\nu = 0.12 \, {\rm eV}$, derived from Planck and BOSS DR12 data \citep{Planck:2018vyg}.

This necessitates a brief review of the role of massive neutrinos on the expansion history of our universe and structure formation. In the early universe, neutrinos contribute to a major fraction of the radiation component, while in the matter domination era (MDE), it adds significantly small fraction ($\sim 1 \%$) to the energy budget of the universe as matter component \citep{Lesgourgues:2013sjj,Lesgourgues:2006nd,Bashinsky:2003tk}. This non-relativistic transition from the relativistic phase leaves imprints on the CMB observables by modifying the expansion history. On large scales, massive neutrinos behave like cold dark matter (CDM) and contribute to structure formation. In contrast, on small scales, their free-streaming nature suppresses the growth of CDM perturbations, leading to a scale-dependent reduction in the matter power spectrum \citep{lesgourgues_mangano_miele_pastor_2013,Dolgov:2002wy,Bashinsky:2003tk}

Recent studies have put a considerable effort towards modelling the non-linear impact of massive neutrinos on large-scale structure (LSS) observables. These include the matter power spectrum \citep{Blas:2014hya,Garny:2020ilv,Garny:2022fsh,Aviles:2020cax,Noriega:2022nhf}, bispectrum \citep{Levi2016tlf,deBelsunce:2018xtd,Racco:2024lbu}, halo bispectrum \citep{Yankelevich:2022mus}, and void statistics \citep{Zhang:2019wtu,Vielzeuf:2023fqw}. Extensive analyses using N-body simulations and the Effective Field Theory (EFT) of  LSS have shown that the bispectrum, especially in redshift-space, acts as a complementary tool to the power spectrum in constraining cosmology, that in turn enhances the sensitivity to the neutrino mass scale \citep{Hahn:2019zob,Hahn:2020lou,Noriega:2024lzo,Ruggeri:2017dda,Villaescusa-Navarro:2017mfx,Castorina:2015bma}. While early bispectrum studies focused primarily on isotropic components (monopole), recent developments highlight the value of accounting for redshift-space anisotropies \citep{Slepian:2016weg,Verdiani:2025znc}, motivating our current investigation.

In this paper, we study the redshift-space bispectrum multipoles $B_L^m(k_1, \mu, t)$, constructed by expanding the bispectrum in spherical harmonics basis. Here, $k_1$ is the largest side of the triangle and $(\mu, t)$ encode the shape of the triangle. Moving beyond earlier analyses that primarily considered lower order multipoles \citep{Ruggeri:2017dda,Hahn:2019zob,Hahn:2020lou},  we compute the full set of neutrino-corrected bispectrum multipoles up to all non-zero $L$ and $m$. Our results demonstrate that massive neutrinos imprint distinctive signatures across a wide variety of triangle configurations, including higher multipoles such as ($L=4, m=0$) and beyond. The impact of neutrinos enters through modifications to the growth factor and second-order perturbation theory kernels. While these effects are small in real space, they are amplified in redshift space due to their interplay with bias and Finger-of-God (FoG) effects. The extent of these corrections varies across triangle shapes and is especially sensitive to the non-linear bias parameter.

To explore the observability of these effects, we perform a forecast of the signal-to-noise ratio (SNR) for bispectrum multipoles relevant to the \textit{Euclid} galaxy survey, which will cover redshifts of roughly $\sim 10^8$ galaxies across $15,000$ square degrees up to $z \sim 2.0$ \citep{Amendola:2016saw,Euclid:2021icp,Euclid:2024vss,Euclid:2024yrr,Euclid:2021icp,Euclid:2021qvm}. Our analysis includes observational effects such as FoG damping and shot noise, and extends the SNR evaluation to multipoles beyond the commonly studied lower order terms. This sets the stage for a detailed investigation of how bispectrum multipoles, especially at higher orders, can serve as sensitive probes of neutrino physics.

The paper is organized as follows. In Sec.~\ref{sec:bispectrum_formalism}, we review the  computation of the bispectrum induced from gravitational non-linearities in Fourier space and redshift-space. Following this, in Sec.~\ref{sec:bispectrum_formalism_with_nu}, we compute the redshift-space corrections to the bispectrum multipoles induced by massive neutrinos, providing quantitative estimates for all multipole moments. Then in Sec.~\ref{sec:formalism_snr}, we present a detailed formalism for computing the signal-to-noise ratio (SNR) in the presence of massive neutrinos. In Sec.~\ref{subsec:snr_prediction}, we provide SNR predictions for various bispectrum multipoles in the context of \textit{Euclid}, accounting for shot noise and Finger-of-God effects. We identify the multipoles and triangle configurations where neutrino-induced corrections are most pronounced. Finally in Sec.~\ref{sec:discussions}, we have outlined the results of our analysis and future prospects.

\section{Formalism}

\subsection{Induced Bispectrum}
\label{sec:bispectrum_formalism}

In cosmological perturbation theory, the evolution of matter fluctuations is usually described by decomposing the total matter density field into a homogeneous background and a perturbation part: $\rho(\mathbf{x},\tau) = \Bar{\rho}(\tau)(1+\delta(\mathbf{x},\tau))$, where  $\Bar{\rho}(\tau)$ is the  background density  that scales as $\tau^{-2}$ during the matter dominated epoch (MDE), and  $\delta (\mathbf{x},\tau)=\delta \rho(\mathbf{x},\tau)/\Bar{\rho}$ is the density contrast. Here, $\tau$ denotes conformal time.  In the linear regime, where $\delta (\mathbf{x},\tau) \ll 1$, different Fourier modes evolve independently, making it convenient to study structure formation in Fourier space.
Under the Einstein-de Sitter (EdS) approximation — which is valid deep inside the MDE, scale and time dependence of density perturbations can be factorized. The density contrast in Fourier space can be expanded as a perturbative series
\citep{Bernardeau:2001qr, dodelson2024modern}:

\begin{eqnarray}
    \delta(\mathbf{k},\tau) = \sum_{n=1}^{\infty} \delta_n(\mathbf{k},\tau)\, ,
    \label{delta_m}
\end{eqnarray}
where $\mathbf{k}$ is the Fourier conjugate of $\mathbf{x}$.
The validity of the EdS approximation in this regime allows for a separation of variables approach, where time evolution and spatial dependence are handled independently \citep{Bernardeau:2001qr,Fasiello:2022lff}.
At each order in perturbation theory, the density contrast and velocity divergence fields are expressed as convolutions over products of linear fields weighted by mode-coupling kernels $F_n$ and $G_n$ respectively \citep{Bernardeau:2001qr,dodelson2024modern,Scoccimarro:1995if,Fasiello:2022lff,Scoccimarro:1999ed,Scoccimarro:2000sn}:
\begin{eqnarray}
    \delta_n(\mathbf{k},\tau) = \int d^3 \mathbf{k_1}\;...\;d^3 \mathbf{k_n} \;\delta_{\mathrm{D}}^{[3]} \left(\mathbf{k}-\sum_{n=1}^{\infty} \mathbf{k_i} \right) \nonumber \\ \times F_n(\mathbf{k_1},...,\mathbf{k_n}) \delta_1(\mathbf{k_1})\; ... \;\delta_1({\mathbf{k_n}}) \,,
    \label{kernelF2}
\end{eqnarray}
\begin{eqnarray}
    \theta_n(\mathbf{k},\tau) = \int d^3 \mathbf{k_1}\;...\;d^3 \mathbf{k_n} \;\delta_{\mathrm{D}}^{[3]} \left(\mathbf{k}-\sum_{n=1}^{\infty} \mathbf{k_i} \right)\nonumber \\ \times  G_n(\mathbf{k_1},...,\mathbf{k_n}) \delta_1(\mathbf{k_1})\; ... \;\delta_1({\mathbf{k_n}}) \, ,
    \label{kernelG2}
\end{eqnarray}
where $\delta_n(\mathbf{k},\tau)$ and $\theta_n(\mathbf{k},\tau)$ are $n^{\rm th}$ order density and velocity perturbations respectively. $F_n(\mathbf{k_1},...,\mathbf{k_n})$ and $G_n(\mathbf{k_1},...,\mathbf{k_n})$ are symmetrized kernels, describing non-linear mode coupling with $F_1 =1, \; G_1 =1$. $\delta_{\mathrm{D}}^{[3]}$ is the Dirac delta function which ensures that the total momentum is conserved.
If one assumes a Gaussian initial condition (which is the simplest possible distribution of cosmic structures), the three-point correlation function (3PCF) of the linear density field vanishes in standard cosmology by Wicks' theorem \citep{Wick:1950ee,Desjacques:2010jw}. However, even though it works quite well at the linear scales,
there is no compelling reason why the distribution has to be perfectly Gaussian all through. In fact,
at late times, gravitational non-linearities can induce a non-zero three-point correlation function or bispectrum, even at the tree level, which can be an interesting probe of perturbations in large scale structure observations.
Herein lies the relevance of the present analysis.

Given the forms of the kernels $F_2$ and $G_2$, 
which describe the second-order perturbative effects, we can express the second-order density contrast, $\delta_2$, in terms of two linear density contrasts $\delta_1$. 
This provides a way to model the non-linear effects and their contribution to the bispectrum.
The explicit forms of the kernels are well known in literature \citep{Bernardeau:2001qr,Scoccimarro:1995if} which follows as,
\begin{eqnarray}
    F_2(\mathbf{k_1},\mathbf{k_2})={5\over 7}+{\mathbf{k_1}\cdot \mathbf{k_2}\over 2}\left({1\over k_1^2}+{1\over k_2^2}\right)+{2\over7}{(\mathbf{k_1}\cdot \mathbf{k_2})^2\over k_1^2 k_2^2}\, ,
\label{eq:F2_kernel}
\end{eqnarray}
\begin{eqnarray}
G_2(\mathbf{k_1},\mathbf{k_2})={3\over 7}+{\mathbf{k_1}\cdot \mathbf{k_2}\over 2}\left({1\over k_1^2}+{1\over k_2^2}\right)+{4\over7}{(\mathbf{k_1}\cdot \mathbf{k_2})^2\over k_1^2 k_2^2} \, .
\label{eq:G2_kernel}
\end{eqnarray}
The matter bispectrum or the 3PCF in Fourier space is defined as,
\begin{eqnarray}
	B(\mathbf{k_1},\mathbf{k_2},\mathbf{k_3})=V^{-1}\, \langle \delta_2(\mathbf{k_1}) \delta_1(\mathbf{k_2}) \delta_1(\mathbf{k_3}) + {\rm cyc}...\ \rangle \,,
	\label{eq:B_def}
\end{eqnarray}
where ``${\rm cyc}...$" represent cyclic permutations. The bispectrum is non-zero whenever any of the density field is of $2^{\rm nd}$ order and the condition for a closed triangle in Fourier space (\textit{i.e.} $\mathbf{k_1}+\mathbf{k_2}+\mathbf{k_3}= \mathbf{0}$) holds. 
 In order to parametrize any triangle in $k$ space, one needs nine parameters that describe the coordinates of each vertex of the triangle. Translational invariance reduces the number of parameters to six, and  rotational invariance about the line-of-sight further reduces the total number to five. In real space
\footnote{Real space refers to the true distribution of matter
in the universe, unaffected by distortions due to peculiar velocities, 
in contrast to redshift-space where observed positions are altered by these velocities along the line-of-sight. The real space is used in the context of  both $\mathbf{x}$ and 
its Fourier conjugate $\mathbf{k}$. We indicate real and redshift-space by superscripts `$r$'
and `$s$' respectively.}, the bispectrum $B^r(\mathbf{k_1},\mathbf{k_2},\mathbf{k_3})$ depends on the shape and size of the triangle and is independent of the orientation of the triangle. Therefore, the bispectrum can be described either by the 
length of the three $\mathbf{k}$ sides of the triangle,
or by lengths of any of those two sides and the angle between them. 
Throughout the paper, we parametrize the size of the triangle by 
$k_1$ with $k_1 = |{\mathbf{k_1}}|$ and the shape dependency is parametrized by 
$t=k_2/k_1$ and
$\mu = \cos{\theta}=-(\mathbf{k_1}\cdot\mathbf{k_2})/ (k_1 k_2)$, where $\theta$ is the angle between  $\mathbf{k_1}$ and $-\mathbf{k_2}$.
We also introduce another ratio, $s=k_3/k_1 =\sqrt{1-2\mu t +t^2}$  in order to simplify analytic expressions.
We further impose a condition $k_1 \geq k_2 \geq k_3$ to identify only the unique triangles.
This sets limits on $\mu$ and $t$, and for a fixed $k_1$, $\mu$ and $t$ vary within
\begin{eqnarray}
0.5 \le t, \, \mu \le 1 \, \, {\rm and}\, \,2 \, \mu \, t \ge 1   \, .  
\end{eqnarray}
In short, a triangle in $\mathbf{k}$ space can be fully characterized by the set $(k_1,\mu,t)$.
For more details on this, the reader can refer to \citet{Bharadwaj:2020wkc}.
In this paper, we fix a baseline value of $k_1=0.2\, \mathrm{h\,Mpc^{-1}}$
for most of the results presented in later sections.
In redshift-space, peculiar velocities introduce anisotropy in the apparent galaxy distribution 
called redshift-space distortion (RSD).
Due to this anisotropy, the bispectrum depends not only on the shape and size of the triangle 
but also on the orientation of the triangle with respect to the line-of-sight (LoS). This orientation can be specified by the cosines of the angles between $\mathbf{k_1}, \mathbf{k_2}, \mathbf{k_3}$  and LoS direction $\hat{\mathbf{n}}$. 
Let us define $\mu_i= (\mathbf{k_i}\cdot\hat{n})/k_i $ where $i=1,2$ and $3$. 
In order to fully quantify the anisotropy of the bispectrum in redshift-space, 
it is necessary to consider triangles of all possible orientations.
We follow the procedure outlined in \citet{Mazumdar:2022ynd} 
and choose a reference triangle in the $x-z$ plane such that,
\begin{eqnarray}
    \mathbf{k_1} = k_1 \hat{z}\,,\\
    \mathbf{k_2} = k_1 t \left[ -\mu\hat{z} + \sqrt{1-\mu^2}\hat{x} \right]\,
\end{eqnarray}
and $\mathbf{k_3}=-(\mathbf{k_1}+\mathbf{k_2})$.
This reference triangle can be rotated to obtain all possible orientations
and the rotations can be parametrized using the Euler angles ($\alpha,\beta,\gamma$) which refer to successive rotations along the $z,y$ and $x$ axes respectively \citep{Bharadwaj:2020wkc,Mazumdar:2020bkm,Mazumdar:2022ynd}. 
Therefore, the bispectrum in redshift-space, denoted by $B^s$,
is a function of six variables :
$(k_,\mu,t,\alpha,\beta,\gamma)$.
Note here that,
$\mu_i$'s are independent of $\alpha$, as the rotation along the $z$ axis coincides with $\hat{\mathbf{n}}$ ($=\hat{\mathbf{z}}$).
So, we have the following coordinate to describe the orientation of the triangle after the Euler rotation,
\begin{eqnarray}
\mu_1 &=& p_z \nonumber \\
\mu_2 &=&-\mu p_z - \sqrt{1-\mu^2} p_x  \\
\mu_3 &=& -(\mu_1 k_1+\mu_2 k_2)/k_3=-s^{-1} [(1-t \mu) p_z + t \sqrt{1-\mu^2} p_x] \nonumber\, ,
\label{eq:coordinate}
\end{eqnarray}
where $p_z =\cos{\beta}$ and $p_x =-\sin{\beta}\cos{\gamma}$ are components of a unit vector $\hat{\mathbf{p}}$.

We adopt the approach outlined in \citet{Scoccimarro:1999ed} and \citet{Mazumdar:2020bkm}
and quantify the redshift-space anisotropy of the bispectrum by expanding it in spherical harmonics with respect to $\hat{\mathbf{p}}$,
\begin{eqnarray}
{B}^{m}_{L}(k_1,\mu,t) =\sqrt{\frac{(2 L +1)}{4 \pi}} \int [Y^m_{L}(\mathbf{\hat{p}})]^* B^s(k_1,\mu,t,\mathbf{\hat{p}}) \, d\Omega_{\mathbf{\hat{p}}} \, ,
\label{eq:bispectrum_sh}
\end{eqnarray}
where $Y_{L}^{m}(\mathbf{\hat{p}})$ are the spherical harmonics covering the $4 \pi$ steradians of the sky $(\beta \in [0, \pi]$, $\gamma \in [0, 2\pi])$. Considering flat-sky approximation \citet{Hamilton:1997zq}, it is possible to show that only the even $L$'s are non-vanishing.
Further, the multipoles are real, obeying the property $B_{L}^{-m} = (-1)^m B_{L} ^{m}$,
and have non-zero values for $0 \leq L \leq 8$ and $0 \leq m \leq 6$.

As mentioned previously, we are interested in the galaxy bispectrum in the presence of redshift-space anisotropies. Galaxies serve as tracers of the underlying matter density field; however, what we observe in redshift-space is the galaxy over-density, not just the matter over-density. The galaxy density contrast is related to the matter density contrast through bias parameters, which encapsulate the complex physics of galaxy formation. These bias models can range from simple linear relations to more intricate non-linear ones, depending on the details of the formation processes. Following \citet{Desjacques:2016bnm}, the galaxy density contrast can be modelled as,
\begin{eqnarray}
    \delta_{\rm g}(k,\tau)=b_1 \delta_{\rm m}(k,\tau)+b_2 (\delta_{\rm m}^2(k,\tau)-\langle \delta(k,\tau)^2 \rangle)+b_{\rm t} S_{\rm m}+... ,
    \label{eq:galaxy_bias}
\end{eqnarray}
where $\delta_{\rm m}$ and $\delta_{\rm g}$ are underlying matter density and galaxy density contrast respectively, $b_1$ is the linear bias and $b_2$ is the non-linear bias parameter. $b_{\rm t}$\footnote{In this analysis, we have not considered the terms involving the tidal bias. We do not expect this contribution to be significant for the low-redshift galaxy population, as supported by the results presented in Appendix~\ref{sec:appendixB}. Following \citet{Gil-Marin:2014sta}, the kernel that generates the tidal bias is simply a rescaling of $F_2(k_1,\mu,t)\rightarrow F_2(k_1,\mu,t)+\frac{2 b_{\rm t}}{3 b_1}P_2(\mu)$, where $P_2$ is the Legendre polynomial of order 2. Also Following \citep{2009JCAP...08..020M,Chan:2012jj,Baldauf:2011bh}, the tidal bias can be written in terms of the linear bias, $b_{\rm t} = - \frac{2}{7}(b_1-1)$.} 
is tidal bias corresponding to the matter tidal tensor $S_{\rm m}$. In general, all these bias parameters are scale- and time-dependent.
Adopting the above-mentioned bias modelling and keeping terms up to $b_2$, the bispectrum in redshift-space can be expressed in terms of $\mu_i$'s \citep{Scoccimarro:1999ed,Slepian:2016weg,Gil-Marin:2014sta} as follows,
\begin{eqnarray}\label{eq:bispectrum_eq1}
B^s(\mathbf{k_1},\mathbf{k_2},\mathbf{k_3}) &=& 2  b_1^{-1}(1+\beta_1 \mu_1^2)(1+\beta_1 \mu_2^2)\Big\{ F_2(\mathbf{k_1},\mathbf{k_2})+{\gamma_2 \over 2}
\nonumber \\ &+& \mu_3^2\beta_1 G_2(\mathbf{k_1},\mathbf{k_2})
-\frac{b_1 \beta_1 \mu_3 k_3}{2} \Big[ {\mu_1\over k_1}(1+\beta_1 \mu_2^2) \nonumber \\ &
+ & {\mu_2\over k_2}(1+\beta_1 \mu_1^2)\Big]\Big\} P(k_1) P(k_2) + 
{\rm cyc ...}\, ,
\end{eqnarray}
where $\mu_i$'s are defined in Eq.~\ref{eq:coordinate} and $\beta_1$ is related to the 
logarithmic growth rate through $\beta_1=f/b_1$ and $\gamma_2 \equiv b_2/b_1$. This formalism sets the stage to account for the effects of massive neutrinos in redshift space.
\subsection{Neutrino-Induced Modifications to the Bispectrum}
\label{sec:bispectrum_formalism_with_nu}

Neutrinos are well-known to induce a characteristic suppression in both CMB and matter power spectra. When relativistic, they contribute nearly $41 \%$ of the total radiation content of the Universe \citep{lesgourgues_mangano_miele_pastor_2013,Lesgourgues:2006nd}.
After transitioning to the non-relativistic regime, neutrinos account for about $\sim 1\%$ of the total matter density.  The cosmological effects of neutrinos—whether in the radiation-dominated or matter-dominated regime—have been widely explored in the literature \citep{Dolgov:2002wy,Bashinsky:2003tk}. In this work, we concentrate on the latter case, specifically the impact of massive neutrinos on the formation of large-scale structures. Massive neutrinos become non-relativistic well inside the matter-dominated epoch. 
The fraction of total neutrino mass over the total matter, $f_\nu$, in this epoch
can be calculated as $f_\nu =\Omega_{\nu}/\Omega_{{\rm m}}$ where 
$\Omega_{{\rm m}}$ is the matter density parameter, and 
$\Omega_{\nu} h^2 = \sum_{i} m_{\nu,i}/93.14\,\mathrm{eV}$ with 
$m_{\nu,i}$ being the mass of different neutrino species and the sum 
runs over all the species.
In the fluid approximation, the evolution of the neutrino velocity divergence, $\theta(\mathbf(k,\tau)$
 in conformal time $\tau$ is governed by the linearized equation of motion,
\begin{eqnarray}
\frac{d\theta(\mathbf{k}, \tau)}{d\tau} + \mathcal{H}(\tau) \theta(\mathbf{k}, \tau) 
+ \left( \frac{3}{2} \mathcal{H}^2(\tau) - k^2 c_s^2(\tau) \right) \delta(\mathbf{k}, \tau) = 0 \, ,
\end{eqnarray}
where $\mathcal{H} \equiv \frac{d \ln a}{d\tau}$ is the conformal Hubble parameter, $\delta(\mathbf{k},\tau)$ is the density contrast, and $c_s(\tau)$ is the effective sound speed of neutrinos, approximated by their thermal velocity dispersion $\sigma_v(\tau)$ \citep{Shoji:2010hm,Levi2016tlf,Pal:2023dcs,Nascimento:2023psl}.
The velocity dispersion introduces a time-varying length scale, analogous to the Jeans scale, known as the free-streaming scale. It characterizes the scale below which neutrinos cannot cluster due to their large thermal velocities. The corresponding comoving free-streaming scale is approximately given by,
\begin{eqnarray}
k_{\mathrm{FS}}(\tau) &\equiv & \sqrt{\frac{3}{2}} \, \frac{\mathcal{H}(\tau)}{c_s(\tau)} 
\simeq \sqrt{\frac{3}{2}} \, \frac{\mathcal{H}(\tau)}{\sigma_{\nu}(\tau)} \nonumber  \\
& \simeq &  \frac{3 \sqrt{a(\tau)\, \Omega_m}}{2} \left( \frac{m_\nu}{\mathrm{eV}} \right) h \, \mathrm{Mpc}^{-1}\, .
\label{kfs}
\end{eqnarray}
For modes with $k \ll k_{\rm FS}$, neutrinos effectively behave as CDM, enhancing the growth of structure. On the other hand, for $k \gg k_{\rm FS}$, the free-streaming motion suppresses the formation of structures by dampening perturbations on small scales.
Analytical studies using the two-fluid framework \citep{Fuhrer:2014zka,Blas:2014hya,Levi2016tlf} show that for $k \gg k_{\rm FS}$, the matter power spectrum and matter bispectrum experience a suppression of 
$\sim (1-8 f_\nu)$ and $\sim (1-13 f_\nu)$ respectively. Similar results have been obtained using the Effective Field Theory (EFT) of Large-Scale Structure (LSS) \citep{Senatore:2017hyk} for the power spectrum, whereas, analyses of the bispectrum in this framework predicts that matter bispectrum is approximately $16 f_\nu$  times dark matter bispectrum on scales $k \gg k_{\rm FS}$ \citep{deBelsunce:2018xtd}.

In this work, we focus on studying the effects of massive neutrinos on the galaxy bispectrum. In particular, we aim to investigate how redshift-space anisotropies in the bispectrum are modified in the presence of massive neutrinos.
Following Eqs.~\ref{eq:bispectrum_sh} and \ref{eq:bispectrum_eq1}, the bispectrum in redshift-space can be expanded in spherical harmonics. We will search for correction to various multipoles in this basis when neutrino perturbation is taken into account.
As pointed out in \citep{Mazumdar:2020bkm,Mazumdar:2022ynd}, in standard $\Lambda$CDM scenario, different multipoles of the bispectrum in redshift-space exhibit different patterns depending on the size and shape of the triangles.

We are now in a position to discuss the effect of massive neutrinos on the perturbation theory (PT) kernels. Considering neutrinos and CDM in a two fluid framework in EdS background, modifications to the growth factor can be calculated as~ 
$D(k,\tau)^{f_\nu \neq 0} = (1-\frac{3}{5}f_\nu)D(k,\tau)^{f_\nu=0}$
\citep{Lesgourgues:2006nd,lesgourgues_mangano_miele_pastor_2013,Wong:2008ws,Hu:1997mj}, where $D(k,\tau)^{f_\nu=0}$ is the growth factor of CDM+baryon fluid. The modified growth factor leads to modified PT kernels in the presence of neutrinos \citep{Wong:2008ws}. We denote the modified density and velocity kernels by $\Tilde{F}$ and $\Tilde{G}$ respectively. The modified kernels can be written as follows,
\begin{eqnarray}
\label{eq: modified_kernel}
\Tilde{{F}}_2(\mathbf{k}_1,\mathbf{k}_2) = \frac{15A_1 + 2A_4}{21} P_0(\mathbf{\hat{k}_1}\cdot \mathbf{\hat{k}_2}) + \frac{1}{2}\left( A_2\frac{k_1}{k_2}+ A_3\frac{k_2}{k_1}\right) \nonumber  \\   \times P_1(\mathbf{\hat{k}_1}\cdot \mathbf{\hat{k}_2}) + \frac{4A_4 }{21}P_2(\mathbf{\hat{k}_1}\cdot \mathbf{\hat{k}_2}),
\end{eqnarray}
\begin{eqnarray}
\Tilde{G}_2(\mathbf{k}_1,\mathbf{k}_2) = \frac{12C_1 + 4C_4}{21} P_0(\mathbf{\hat{k}_1}\cdot \mathbf{\hat{k}_2}) 
+ \frac{1}{2}\left( C_2\frac{k_1}{k_2}+C_3\frac{k_2}{k_1}\right) \nonumber \\ P_1(\mathbf{\hat{k}_1}\cdot \mathbf{\hat{k}_2}) + \frac{8C_4 }{21}P_2(\mathbf{\hat{k}_1}\cdot \mathbf{\hat{k}_2}).
\end{eqnarray}
where $P_{\ell}$'s are Legendre polynomial of order $\ell$. The $A$ and $C$ coefficients are defined as follows \cite{Wong:2008ws},
\begin{eqnarray}
    A_1 &=&\frac{7}{10}\sigma_{11}^{(2)}(k_1,k_2)[f(k_1)+f(k_2)], \nonumber \\
    A_2&=&f(k_2)[\sigma_{11}^{(2)}(k_1,k_2)+\sigma_{12}^{(2)}(k_1,k_2)f(k_2)], \nonumber\\
    A_3&=&f(k_1)[\sigma_{11}^{(2)}(k_1,k_2)+\sigma_{12}^{(2)}(k_1,k_2)f(k_1)], \nonumber \\
    A_4&=&\frac{7}{2}\sigma_{12}^{(2)}(k_1,k_2)f(k_1)f(k_2), \nonumber \\
    C_1&=&\frac{7}{6}\sigma_{21}^{(2)}(k_1,k_2)[f(k_1)+f(k_2)],\nonumber \\
    C_2&=&f(k_2)[\sigma_{11}^{(2)}(k_1,k_2)+\sigma_{12}^{(2)}(k_1,k_2)f(k_2)],\nonumber \\
    C_3&=&f(k_1)[\sigma_{11}^{(2)}(k_1,k_2)+\sigma_{12}^{(2)}(k_1,k_2)f(k_1)],\nonumber\\
    C_4&=&\frac{7}{4}\sigma_{22}^{(2)}(k_1,k_2)f(k_1)f(k_2),
\end{eqnarray}
and $f(k_i)=\partial \ln{D(k_i,\tau)^{f_\nu \neq 0}}/ \mathcal{H} \partial \tau$ is the logarithmic growth rate in presence of neutrinos \citep{Lesgourgues:2006nd,lesgourgues_mangano_miele_pastor_2013}. 
Although in $\Lambda$CDM background with massive neutrinos, the growth rate is both scale and time dependent \citep{Nascimento:2023psl, Kamalinejad:2022yyl, Pal:2023dcs}, the EdS approximation is adequate for the redshift range of our interest. The scale and time dependent behaviour of the linear kernel $\sigma^{(2)}$ has been calculated in \cite{Wong:2008ws},
\begin{eqnarray}
\label{eq:nu_corrected_matrix}
&\sigma^{(2)}(k_1,k_2)=\frac{1}{\mathcal{N}^{(2)}}
    \begin{bmatrix}
                2\omega^{(2)}+1 &2\\
                3(1-f_{\nu})&2\omega^{(2)}
    \end{bmatrix},\\
    &\omega^{(2)}(k_1,k_2)=f(k_1)+f(k_2),\nonumber\\
    &\mathcal{N}^{(2)}=(2\omega^{(2)}+3)(\omega^{(2)}-1)+3f_{\nu}. \nonumber
\end{eqnarray}
Given the growth rate $D(k,\tau)$, we can obtain $f$
and the matrix $\sigma$ following the computation of $\mathcal{N}^{(2)}$ and $\omega^{(2)}$. This in turn allows to compute the $A_i$'s and $C_i$'s from which we can obtain neutrino corrected $\Tilde{F}_2$  and $\Tilde{G}_2$.

Using the modified growth factor, $\Tilde{F_2}$ can be written as,
\begin{eqnarray}
\Tilde{F}_{12}(\mu,t) &=& \frac{1}{14} \left(4 \mu ^2-7 \mu  t-\frac{7 \mu }{t}+10\right)-\frac{6 f_\nu}{245} (-1+\mu^2)\, ,\nonumber\\
\Tilde{F}_{23}(\mu,t)&=&\frac{7 \mu +\left(3-10 \mu ^2\right) t}{14 t s^2}-\frac{6f_\nu}{245 s^2} (-1+(\mu-t)^2/s^2)\, ,\\
\Tilde{F}_{31}(\mu,t)&=&\frac{t^2 \left(-10 \mu ^2+7 \mu  t+3\right)}{14 s^2}-\frac{6f_\nu}{245 s^2} (-1+(\mu t-1)^2/s^2) \nonumber \,.
\label{eq:modified_kernelF2}
\end{eqnarray}
Here $\Tilde{F}_{ij}$ represents $\Tilde{F_2}(\mathbf{k_i},\mathbf{k_j})$. Note that we only consider correction up to first order in $f_\nu$. 
Similarly, the modified $G_2$ kernels can be written as follows,
\begin{eqnarray}
\Tilde{G}_{12}(\mu,t) &=& \frac{1}{14} \left(8 \mu ^2-7 \mu  t-\frac{7 \mu }{t}+6\right) \nonumber \\ &+& f_\nu \frac{3}{490}\left(-34+49(t+1/t)\mu-64 \mu^2 \right) \, , \\
\Tilde{G}_{23}(\mu,t)&=&-\frac{ 6 \mu ^2 t+t-7 \mu}{14 s^2 t}-f_\nu \frac{3}{490 s^2 t}(-49\mu+t(15+34\mu^2)) \, , \nonumber \\
\Tilde{G}_{31}(\mu,t)&=&\frac{t^2 \left(-6 \mu ^2+7 \mu  t-1\right)}{14 s^2} +f_\nu \frac{3 t^2}{490 s^2}(15-49 \mu t+34\mu^2) \nonumber \, .
\label{eq:modified_kernelG2}
\end{eqnarray}
Here also $\Tilde{G}_{ij}$ denotes $\Tilde{G_2}(\mathbf{k_i},\mathbf{k_j})$. 
As previously pointed out by \citet{Kamalinejad:2020izi}, the density kernels have corrections in the coefficients of Legendre polynomial $P_0$ and $P_2$. However, this is not generically true for $\Tilde{F}_{23}$ and $\Tilde{F}_{31}$ terms, where there are corrections to $P_1$ also.
In each $\Tilde{G}_{ij}$ terms, all the Legendre polynomials are affected by the neutrino mass. In \citet{Kamalinejad:2020izi}, the authors examined the isotropic bispectrum by averaging over all possible triangle orientations with respect to the line-of-sight (LoS), effectively isolating only the monopole component. They then performed an expansion in Legendre polynomials of $\mu$, identifying a non-degenerate neutrino signature in the dipole moment. However, their analysis was limited to pre-cyclic label computations. In contrast, our work focuses on analyzing all multipole moments by expanding the bispectrum in a spherical harmonic basis.

In what follows, we compute the redshift-space bispectrum, as defined in Eq.~\ref{eq:bispectrum_eq1}, by employing the modified perturbation theory kernels presented in Eqs.~\ref{eq:modified_kernelF2} and \ref{eq:modified_kernelG2}. Besides incorporating the modified kernels, we also incorporate the modification of the growth factor $\beta_1$ in Eq.~\ref{eq:bispectrum_eq1}. Once the complete expression for the tree-level bispectrum is obtained, we proceed to expand it in the spherical harmonic basis, following the formalism outlined in Eq.~\ref{eq:bispectrum_sh}. This decomposition allows us to isolate and analyze individual multipole moments, providing a clearer understanding of the anisotropic signatures introduced by neutrinos.

In order to isolate the explicit corrections to the induced anisotropy in each multipole arising from the presence of massive neutrinos, we compute the tree-level bispectrum using matter power spectra obtained from the Boltzmann code \texttt{CLASS} \citep{2011arXiv1104.2932L,2011JCAP...07..034B}, while deliberately excluding the effects of neutrino perturbations in the matter power spectrum. This allows us to trace the effects of neutrinos in individual multipoles due to redshift-space anisotropy. However, it is necessary to analyze the real space configuration first, as presented in the following section. 

\subsubsection{Real Space corrections}
\label{subsec:real_space_bispectrum_with_nu}

\begin{figure}
    \centering
    \includegraphics[width=0.8\linewidth]{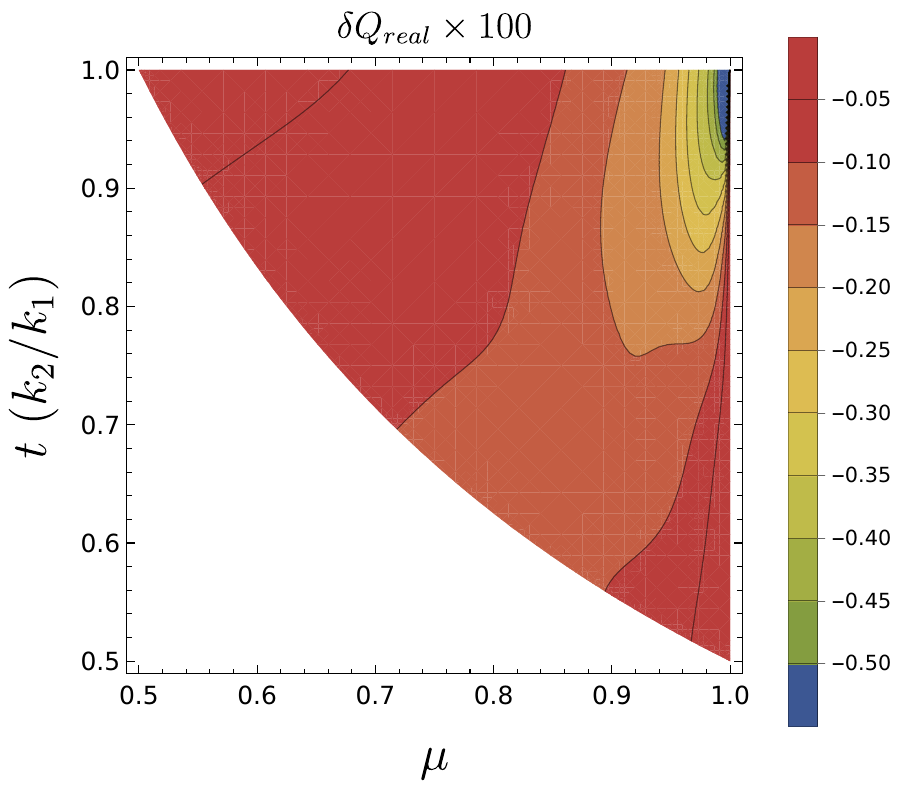}
    \caption{Difference in the real-space reduced bispectrum, 
    $\delta Q_{\rm real}=Q_{\rm real}^{f_{\nu}= 0}-Q_{\rm real}^{f_{\nu}\neq 0}$, induced by massive neutrinos with total mass $\sum m_\nu =0.12 \, {\rm eV}$, shown across the $(\mu,t)$ plane. 
    We fix $k_1 =0.2\, {\rm Mpc}^{-1}$, linear bias $b_1=1.18$, and non-linear bias $\gamma_2=-0.9$. The matter power spectrum is evaluated at redshift $z = 0.7$ using \texttt{CLASS}. 
    Here, the correction remains small across all triangle configurations, typically below 
    $0.1\%$, consistent with the mild impact of neutrinos in real space. Maximum differences occur near
    the squeezed limit.
    }
    \label{fig:real_space_B_correction}
\end{figure}

Before discussing the redshift-space correction, let us first clarify why such corrections are of interest. In real space, the tree level bispectrum can be expressed in the following way,
\begin{eqnarray}
B^r(\mathbf{k_1},\mathbf{k_2},\mathbf{k_3}) = 2 b_1^{-1}\left[\Tilde{F_2}(\mathbf{k_1},\mathbf{k_2})+{\gamma_2 \over 2}\right]P^r(k_1) P^r(k_2) \nonumber \\ + {\rm cyc ...}\, ,
\label{eq:real_B}
\end{eqnarray}
where $\Tilde{F_2}$ and $\Tilde{G_2}$ are modified second-order kernels in presence of neutrinos and $\gamma_2$ ($=b_2/b_1$) denotes the non-linear bias; and $P^r(k)$ is the 
real space matter power spectra. We do not include the neutrino corrections in the matter power spectrum \footnote{The matter power spectrum $P^r(k)$ is computed using \texttt{CLASS} \cite{2011arXiv1104.2932L,2011JCAP...07..034B} with the following fixed cosmological parameters: $\omega_{\rm cdm} =0.1201,\, \omega_{\rm b}=0.02238,\, {\rm h}=0.6781,\, A_{s}=2.1\times 10^{-9},\, n_s=0.9660, \,\tau_{reio}=0.0543$. } for the study in the current section. 
We define the reduced bispectrum in real space as \citep{Bharadwaj:2020wkc},
\begin{eqnarray}
    Q^r(k_1,\mu,t)= {b_1 \, B^r(k_1,\mu,t)\over 3 [P^{r}(k_1)]^2}\, . 
\label{eq:real_Q}
\end{eqnarray}
As mentioned earlier, neutrinos not only suppress the linear growth but also modify the second-order perturbation kernels. This leads to two distinct effects: (i) a suppression in the tree-level bispectrum of order $\sim 13 f_\nu$ due to the change in the linear density contrast coming from the free-streaming nature of neutrinos, and (ii) an additional correction arising from the modification of the non-linear kernels. The first effect has been explored using both linear and non-linear perturbation theory in  \citep{Fuhrer:2014zka,Levi2016tlf,Garny:2020ilv,Garny:2022fsh}. 
Before presenting the results shown in Fig.~\ref{fig:real_space_B_correction}, let us describe the representation of different triangular configuration in ($\mu-t$) plane used in the figure.
These triangle shapes have been discussed in detail in \citet{Mazumdar:2022ynd,Mazumdar:2020bkm}. Nevertheless, we briefly summarize them here. The top-left corner of Fig.~\ref{fig:real_space_B_correction} ($\mu \rightarrow 1/2,t \rightarrow 1$) corresponds to equilateral triangles. The top-right corner ($\mu \rightarrow 1, t \rightarrow 1$) represents the squeezed triangles, where $\mathbf{k_1}=-\mathbf{k_2}$ and $\mathbf{k_3}\rightarrow \mathbf{0}$. Further, the bottom-right corner ($\mu \rightarrow 1, t \rightarrow 1/2$) corresponds to stretched triangles with $\mathbf{k_2}=\mathbf{k_3}=-\mathbf{k_1}/2$. The right boundary (\textit{i.e.} $\mu=1$) configures linear triangles with $\mathbf{k_1}, -\mathbf{k_2}$ and $-\mathbf{k_3}$ all being parallel. Similarly, the top boundary ($t=1$) represents the L-isosceles triangles where the two larger sides ($\mathbf{k_1}$ and $\mathbf{k_2}$) are of equal length and the boundary at bottom ($2\mu t =1$) represents S-isosceles triangles where the smaller sides ($\mathbf{k_2}$ and $\mathbf{k_3}$) are of equal length. Finally, the diagonal line ($\mu=t$) corresponds to the right-angle triangles, whereas the upper and lower half planes with ($\mu >t$) and ($\mu <t$) correspond to the acute and obtuse triangles, respectively.

In Fig.~\ref{fig:real_space_B_correction} we present the difference in $Q^r$, namely, $\delta Q_{\rm real}$ in the ($\mu-t$) space, coming from the modification of the PT kernels only, in presence of neutrinos with $\sum m_\nu=0.12\, {\rm eV}$. The percentage difference in $Q^r$ is defined as,
$\Delta Q_{\rm real}=(\delta Q_{\rm real} \times 100)/Q_{\rm real}^{f_{\nu}= 0}$, where $\delta Q_{\rm real}=Q_{\rm real}^{f_{\nu}= 0}-Q_{\rm real}^{f_{\nu}\neq 0}$.
Here, the superscript $f_{\nu}\neq 0$ refers to model that include massive neutrinos, 
while $f_{\nu} = 0$ denotes model without massive neutrinos.
Here we do not explicitly show the ($\mu,t$) dependence of $Q^r$ to keep the discussion
brief, and only highlight key features. $Q^r$ in the standard scenario, i.e., without the massive neutrinos, has been extensively discussed in \citep{Mazumdar:2020bkm}. There we see that
$Q^r$ is minimum for the equilateral configuration, and the value increases as we
move towards linear triangles. 
$Q^r$ in the presence of massive neutrinos also exhibit similar features. 
Considering the equilateral configuration, $Q^r$ in the presence of massive neutrinos
can be expressed as, 
\begin{eqnarray}
Q^{ r}(1/2,1)=\left(\frac{4}{7}+\frac{9}{245} f_\nu +\gamma_2 \right) \, ,
\label{eq:equilateral_limit}
\end{eqnarray}
which recovers the standard expression in the limit $f_\nu \to 0$.
It is evident from Eq.~\ref{eq:equilateral_limit} that the neutrino-induced corrections are degenerate with the non-linear bias parameter. Considering $\gamma_2=0$, the correction is very small, with $\Delta Q \approx -0.06 \%$, which is also evident from the figure. 
However, assuming a negative non-linear bias $\gamma_2=-0.9$, the correction increases to $\sim 0.1 \%$ for $f_\nu=0.01$. Similarly, in the squeezed limit, $Q^r$ reduced to, $Q^{r}(1,1)=\left(\frac{12}{245} f_\nu +\gamma_2 \right)$, and  the percentage difference compared to the standard scenario again remains negligible, which is apparent from the figure. This motivates us to explore the redshift-space bispectrum, where the effects of neutrinos are expected to be more significant. The details are discussed in the following sections.

\subsubsection{Redshift-space corrections}
\label{subsec:redshift_space_bispectrum_with_nu}

Let us focus on all non-zero multipoles of the redshift-space bispectrum. Instead of explicitly handling the kernel functions as done in \cite{Mazumdar:2020bkm}, we adopt a more straightforward approach. Our analysis relies on  expanding the redshift-space bispectrum, incorporating neutrino-corrected perturbation theory kernels and growth rate, in terms of spherical harmonics. In more concrete words, we start with Eq.~\ref{eq:bispectrum_eq1}, replacing $F_2$ and $G_2$ with $\tilde{F}_2$ and $\tilde{G}_2$, alongside the modified growth factor, and employ Eq.~\ref{eq:bispectrum_sh} to express them in terms of multipoles. 
For further details, including the explicit neutrino corrections to each non-zero multipole, we refer the readers to a GitHub repository, link of which is provided in the `Data Availability' section.  
As previously discussed, realistic predictions with actual galaxy surveys require the inclusion of small-scale effects such as Finger-of-God (FoG) damping and shot noise. However, this comes with a price. With these realistic scenarios, the analytical framework looks less tractable than usual. So, whenever these effects are considered, we rely on numerical methods to assess their impact and to forecast on specific galaxy surveys.

In analogy with the real-space case, we define the dimensionless reduced bispectrum in redshift-space as follows,
\begin{eqnarray}
    Q_L^m (k_1,\mu,t)= {b_1  B_L^m (k_1,\mu,t)\over 3 [P(k_1)]^2}\, . 
\label{eq:rsd_Q}
\end{eqnarray}
In the above,
the bispectrum multipoles have been normalized  using the power spectrum evaluated at the corresponding redshift. While our results are broadly consistent with those in \citet{Mazumdar:2020bkm, Mazumdar:2022ynd}, a key distinction is our choice of normalization, which reflects the redshift-dependence of the power spectrum. This modification does not alter the qualitative features of the bispectrum plots in the absence of massive neutrinos, aside from differences in overall amplitude.

In the present section, we outline our results for redshift $z=0.7$. The corresponding power spectrum are extracted from \texttt{CLASS} with the cosmological parameters as mentioned in the footnote of Sec.~\ref{subsec:real_space_bispectrum_with_nu}. Further, we adopt a linear bias of $b_1 = 1.18$ and a non-linear bias of $\gamma_2 = -0.9$, following \cite{Yankelevich:2018uaz}. The total neutrino mass is taken to  be $\sum m_\nu = 0.12\, {\rm eV}$, corresponding to a fractional mass contribution of $f_{\nu} = 0.01$. Throughout the analysis, we fix $k_1$ to $0.2 \, {\rm Mpc}^{-1}$.
We are interested in quantifying the difference in the reduced bispectrum multipoles as defined in Eq.~\ref{eq:rsd_Q} due to the modification to RSD anisotropy in presence of neutrinos.  This has been materialized by introducing a  variable, $\delta Q_L^m$, which is defined as $\delta Q_L^m={Q_L^m |}_{f_\nu=0}-{Q_L^m |}_{f_\nu \neq 0}$.

\begin{figure*}
    \centering
    \subfloat{\includegraphics[width=0.55\columnwidth]{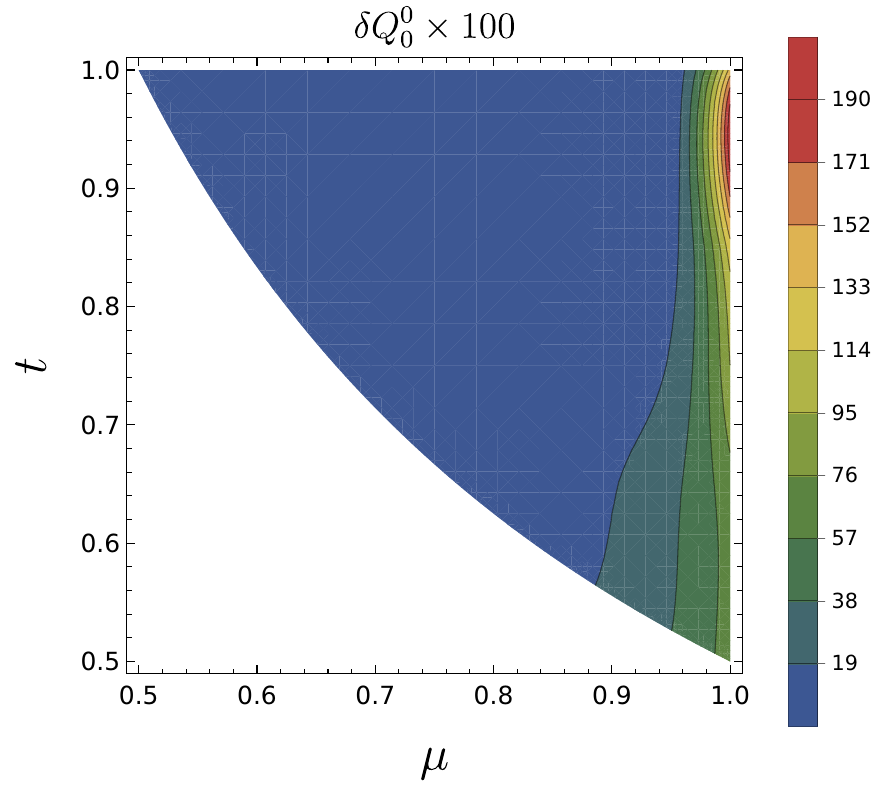}}
    \subfloat{\includegraphics[width=0.55\columnwidth]{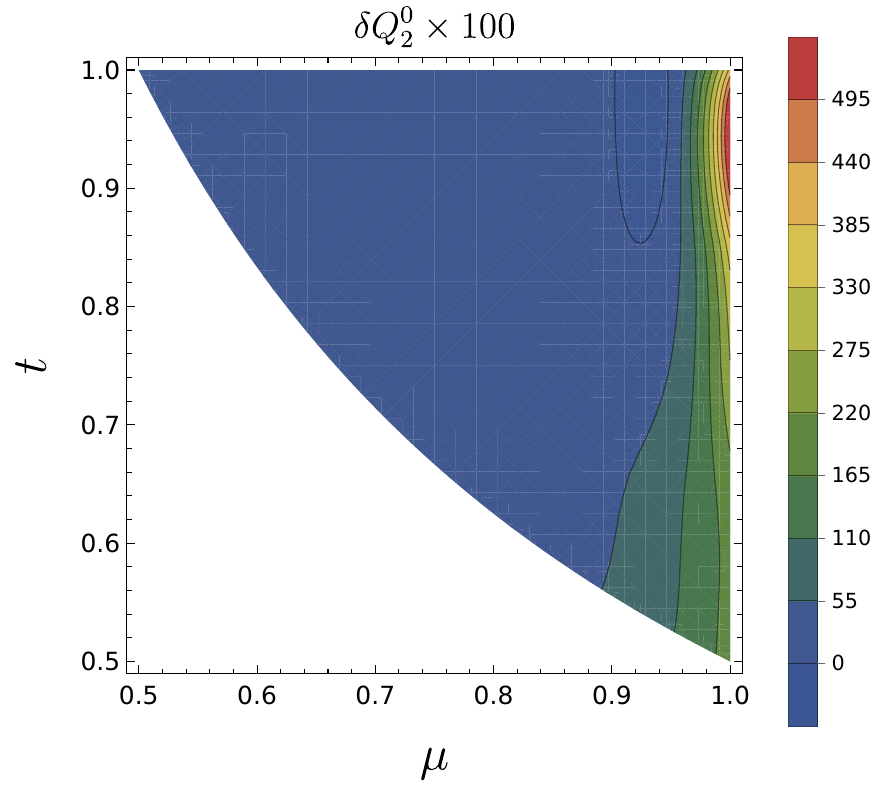}}
    \subfloat{\includegraphics[width=0.55\columnwidth]{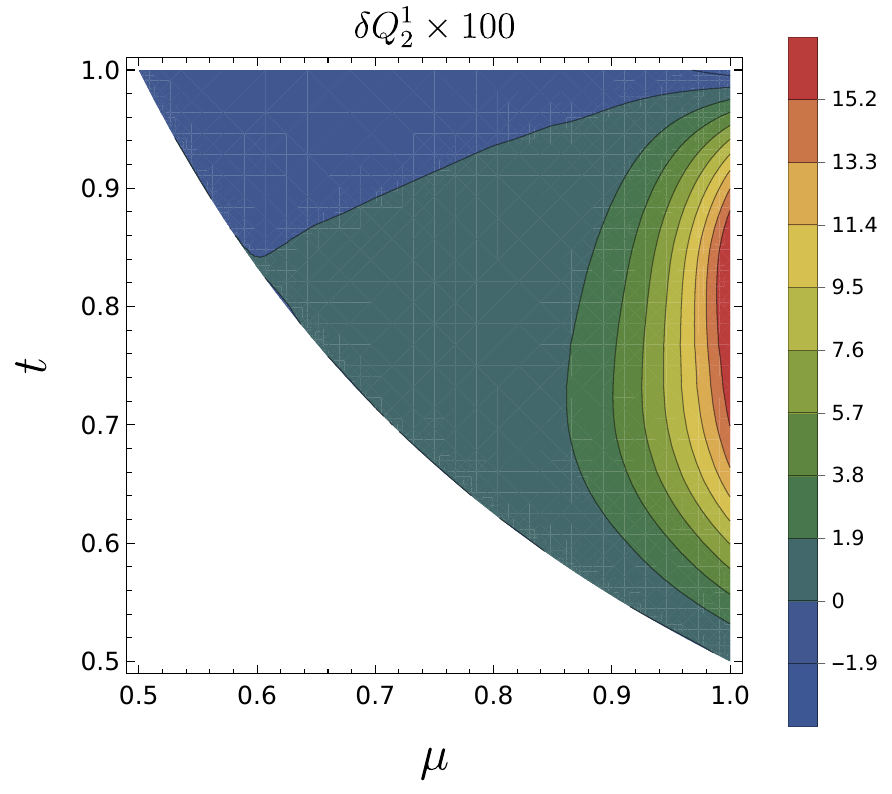}}
    \quad
    \subfloat{\includegraphics[width=0.55\columnwidth]{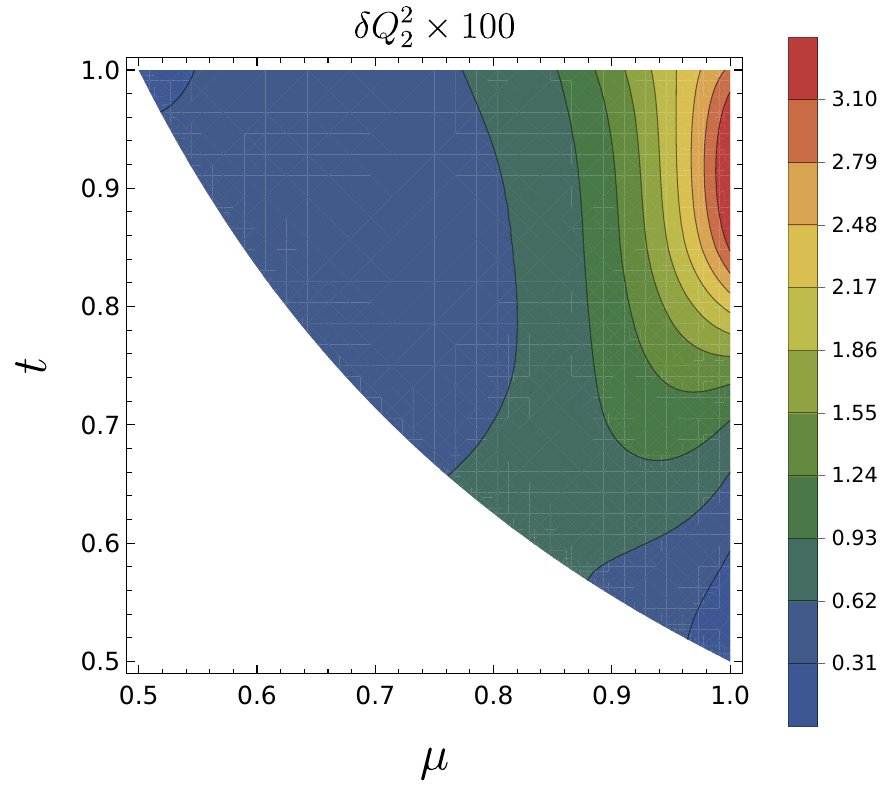}}
    \subfloat{\includegraphics[width=0.55\columnwidth]{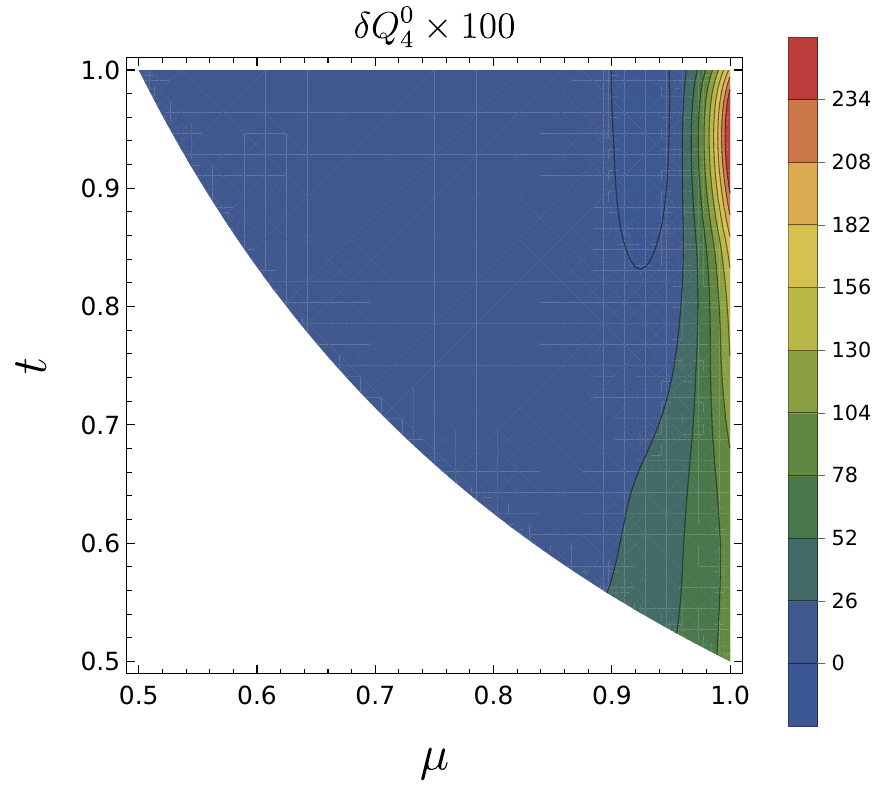}}
    \subfloat{\includegraphics[width=0.55\columnwidth]{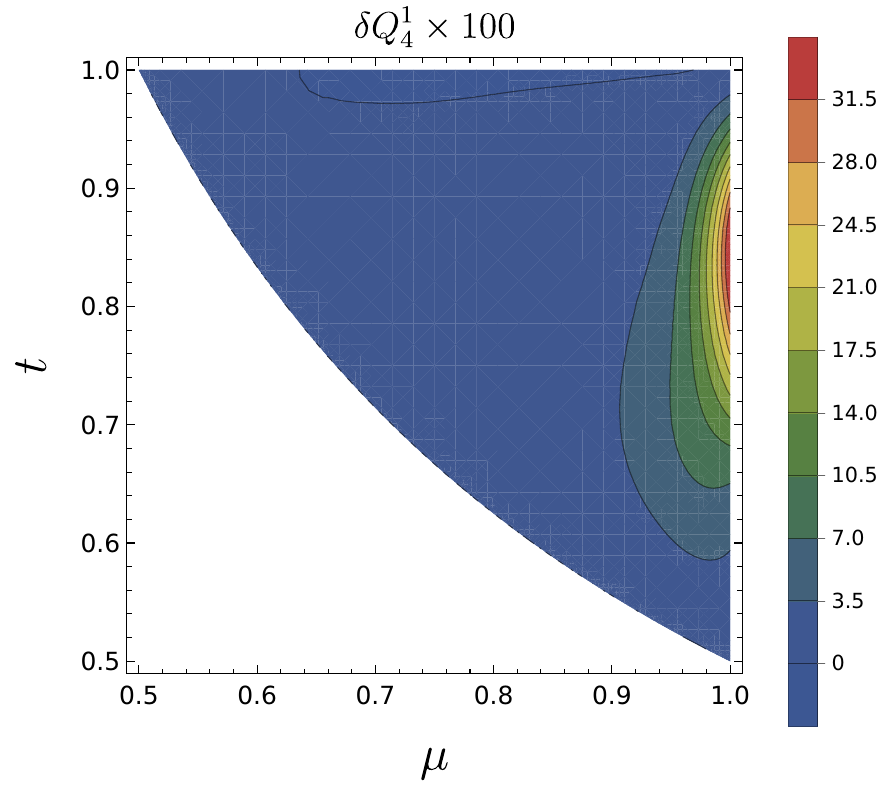}}
    \caption{Fractional difference $\delta Q_L^m$ in the reduced bispectrum multipoles (for $L \leq 4, m \leq 2$) induced by massive neutrinos ($\sum m_\nu = 0.12\,\mathrm{eV}$), plotted over the $(\mu, t)$ plane. Fixed parameters are $k_1 = 0.2\,\mathrm{Mpc}^{-1}$, $b_1 = 1.18$, $\gamma_2 = -0.9$, and $z = 0.7$. Compared to real-space, neutrino effects are significantly amplified in redshift-space. The largest deviations occur near linear triangle configurations ($\mu\rightarrow1$), with corrections reaching the percent level, suggesting strong anisotropic signatures.
    }
    \label{fig:qlm_1}
\end{figure*}

Figure~\ref{fig:qlm_1} (and also Fig.~\ref{fig:qlm_2}) illustrate the variation in $Q_L^m$ due to massive neutrinos. As previously mentioned, only even-$L$ 
multipoles with $L\leq8$ are non-zero and results up to $L=4$ are presented in the main text. We briefly discuss the higher-order multipoles in Appendix~\ref{sec:appendixA}, while their corresponding analytical expressions are provided in an external Github repository, link of which is 
provided in the `Data Availability' section.
As discussed in Sec.~\ref{subsec:real_space_bispectrum_with_nu}, the effect of neutrino mass on the real space bispectrum remains below $0.1\%$ level across all triangle configurations. In contrast, the redshift-space bispectrum multipoles exhibit substantially larger deviations due to the anisotropies inherent in redshift-space distortions. Among the non-zero multipoles, $Q_0^0$, $Q_2^0$, and $Q_4^0$ display qualitatively similar structures.  In each case, the maximum deviation occurs near $\mu \to 1$ and $t \approx 0.94$, corresponding to $k_1 \approx k_2$ and $k_3 = k_{\rm eq}$, where $k_{\rm eq}$ is the scale at matter-radiation equality. 

\begin{figure*}
    \centering
    \subfloat{\includegraphics[width=0.50\columnwidth]{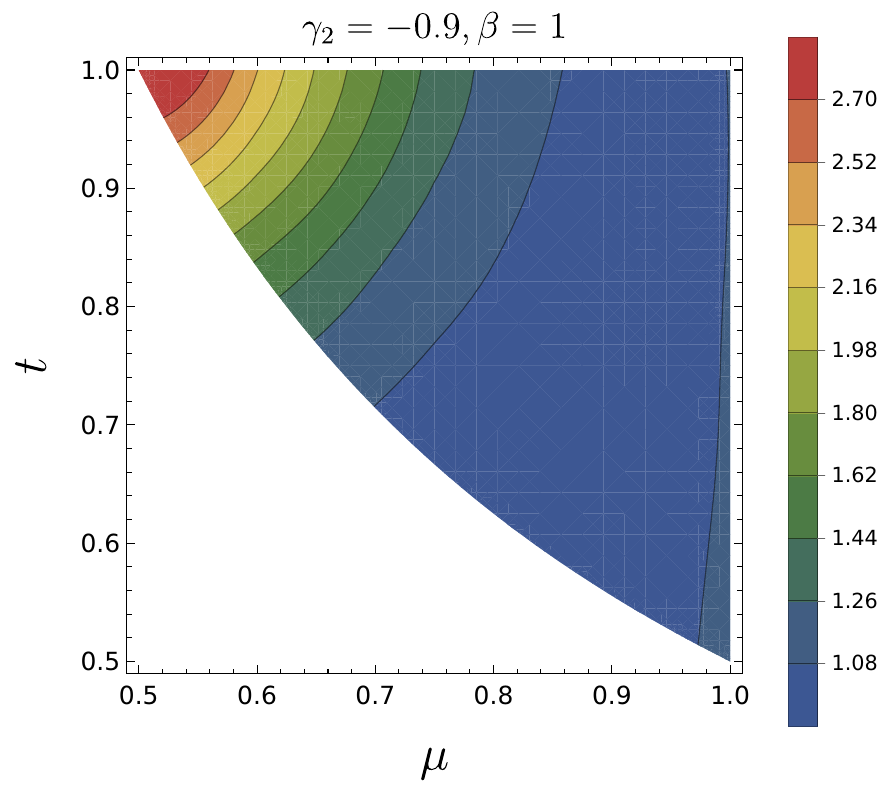}}
    \subfloat{\includegraphics[width=0.50\columnwidth]{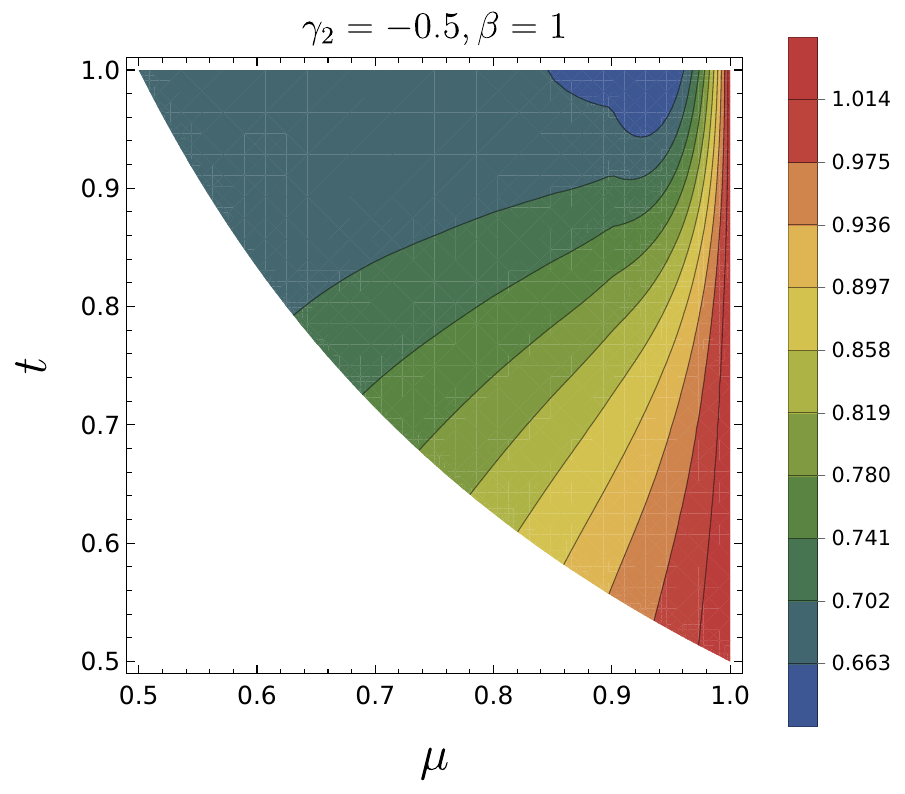}}
    \subfloat{\includegraphics[width=0.50\columnwidth]{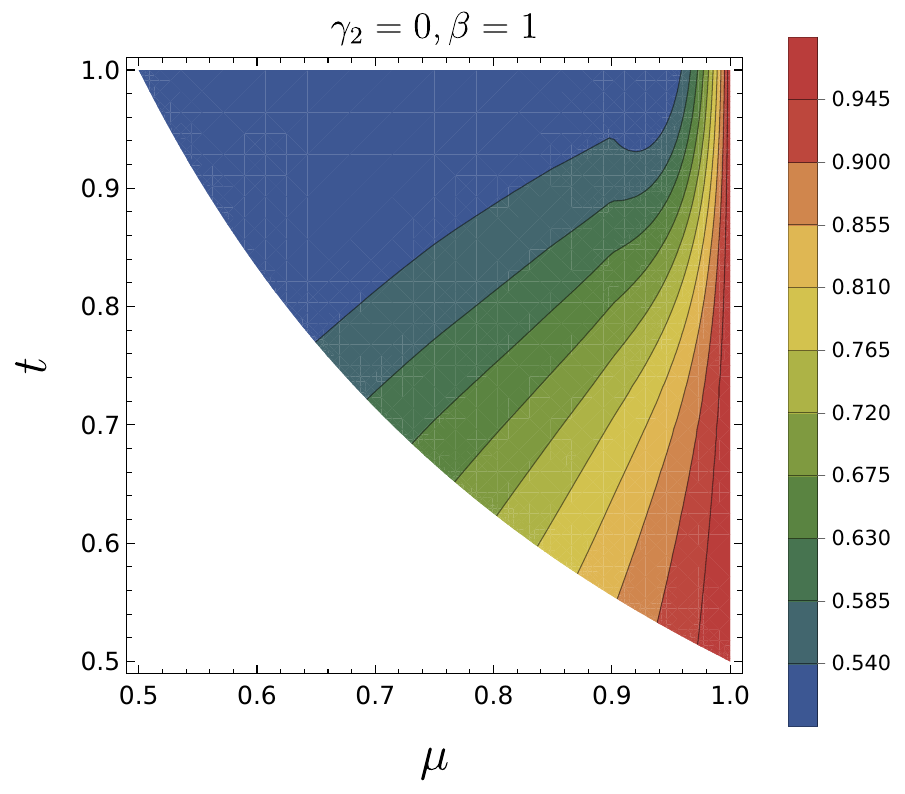}}
    \subfloat{\includegraphics[width=0.50\columnwidth]{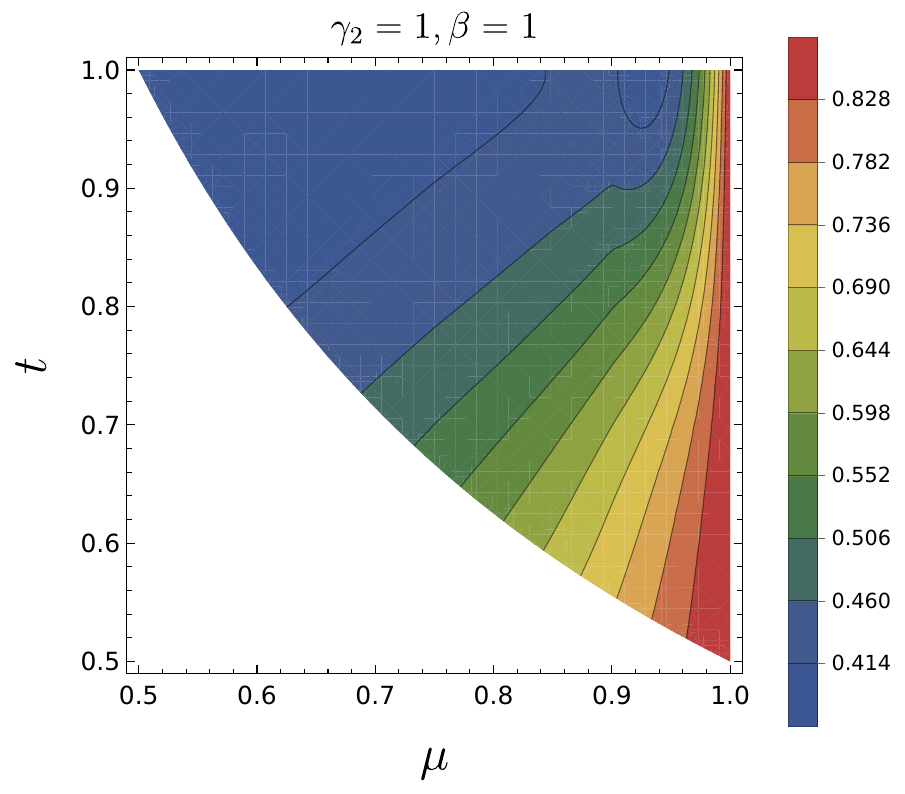}}
    \caption{The percentage difference in the monopole ($Q^0_0$) due to the presence of neutrinos with $\sum m_\nu=0.12 \, {\rm eV}$, varying non-linear bias parameter $\gamma_2$ keeping $\beta_1$ fixed. It is evident from the plots that, the maximum percentage difference decreases with increasing $\gamma_2$ and shifts from equilateral to linear triangle configurations. }
    \label{fig:varying_b2}
\end{figure*}

In absence of neutrinos, the enhancement of $Q_0^0$ due to RSD anisotropy in the equilateral limit is found to be $\sim 2.8$ times $Q^r$, whereas in the linear configuration $Q_0^0 \approx 4.2 \, Q^r$. We do not show the absolute values of $Q_L^m$ here, as they have been thoroughly discussed in \citep{Mazumdar:2020bkm}.
Instead, we focus on the differences in the multipoles ($\delta Q_L^m$) induced by neutrinos. At ($\mu \to 1, t \approx 0.94$) in $Q_0^0$ multipole, we find  
a fractional change of approximately $1.1\%$ due to neutrino mass $\sum m_\nu=0.12\, {\rm eV}$. Although the percentage difference, defined as $(\delta Q_L^m \times 100)/{Q_L^m |}_{f_\nu =0}$), peaks in the equilateral limit at ($\mu = 1/2$, $t \to 1$) with a residual of $\sim 2.8 \%$, this corresponds to a region where the absolute value of the bispectrum in standard scenario is itself small and thus harder to observe. However, we find that, this is a consequence of choosing negative non-linear bias.
It is important to note that the absolute difference, rather than the percentage change, is what matters in our analysis, and we shall see that linear triangle configurations are more promising from a signal-to-noise perspective.
The largest anisotropy in $Q_2^0$ in the standard scenario occurs at linear triangles irrespective of the value of $t$ (where $Q_2^0 \approx 2 \, Q_0^0$). The pattern in $Q_4^0$ is almost similar to $Q_2^0$ except that it is subdominant ($Q_4^0 \approx 0.7 \, Q_0^0)$). The maximum difference in $Q_2^0$ in presence of neutrinos as presented in Fig.~\ref{fig:qlm_1} correspond to $\sim 1.41\%$
at ($\mu \to 1$, $t \approx 0.94$). Similarly, the maximum difference in the presence of neutrinos in $Q_4^0$ multipole occurs at the same position with a residual of $1.74\%$. 
These findings indicate that the most significant impact of massive neutrinos appears in squeezed triangle configurations, though the impact extends along the entire linear triangle branch. The imprints grow more pronounced with increasing multipole order. While the detectability of these higher-order multipoles presents a challenge, it is a key aspect of interest, and we explore it further in the following sections.

In the absence of neutrinos, $Q_2^1$ and $Q_2^2$ multipoles peak in the squeezed limit. At ($\mu \approx 0.9994, t \approx 0.954$), $Q_2^1$ is found to be approximately 0.3 times $Q_0^0$, whereas $Q_2^2$ is more subdominant with $Q_2^2 \approx 0.08 \, Q_0^0$. In the presence of massive neutrinos, 
deviation in $Q_2^1$ from the standard scenario is  found to be $\sim 1.4\%$ at the squeezed limit. For $Q_2^2$, the maximum difference occurs at ($\mu \approx 0.999$, $t \approx 0.97$) with corresponding percentage change $\sim 1.04 \%$. From Fig.~\ref{fig:qlm_1}, we further observe that the $Q_4^1$ multipole exhibits its maximum deviation due to massive neutrinos in configurations close to linear triangles. The peak occurs at ($\mu \approx 0.9995$, $t \approx 0.95168$). In standard scenario, $Q_4^1 \approx 0.21 \, Q_0^0$ at this position. Due to the presence of neutrinos, we obtain a residual change in the multipole of $\sim 1.74 \%$. And finally, the $Q_6^0$ multipole follows a similar trend to $Q_0^0$ and $Q_2^0$, with its maximum located at ($\mu = 1$, $t \approx 0.938$). At this point, we find a fractional change of approximately $2.07\%$ for a total neutrino mass of $0.12\, {\rm eV}$. This pattern persists in higher-order multipoles, although the magnitude of enhancement due to anisotropies decreases significantly with increasing $L$. The corresponding plots are presented in Appendix~\ref{sec:appendixA}.  

So far the results are presented with $\gamma_2 =-0.9$. We also highlight that the bias parameter $\gamma_2$ plays a key role in shaping the multipole response. In Fig.~\ref{fig:varying_b2}, we show how the fractional change in $Q_0^0$ varies with different values of $\gamma_2$, keeping the growth rate parameter $\beta_1 = 1$ fixed. we present the residual percentage change in $Q_0^0$ for $\gamma_2 = -0.5$, $0$, and $1$. When $\gamma_2=-0.9$, the residual percentage change due to the presence of neutrinos peaks in the equilateral configuration. However, due to the small baseline value of the bispectrum in this regime, detectability remains challenging. Conversely, in the squeezed configuration, the effect is smaller but still exceeds $1\%$, making it potentially observable in galaxy surveys such as \textit{Euclid}. We discuss the detection prospects in detail in the following sections. Further, as we increase $\gamma_2$, the dominant corrections shift toward linear triangle configurations, with the strongest enhancements observed in the squeezed limit.  
It is evident from left to right in the figure that increasing $\gamma_2$ leads to a decreasing trend in the fractional differences. Since similar behaviour is observed across other multipoles, we refrain from presenting them all here.

From the above analysis, it can be readily found that,
across all multipoles considered, the influence of neutrinos is most pronounced in the linear triangle configurations. However, particularly intriguing behaviour emerges for  $L \geq 6$ multipoles. Although we do not discuss these in detail here, as their detection is unlikely in the near future.
Interestingly, we find that the relative impact of neutrinos tends to be more significant in higher-order multipoles compared to the lower ones. If future surveys can reliably measure these higher multipoles, it would greatly enhance our ability to detect and constrain neutrino properties. A comprehensive analysis of the bispectrum, including contributions from all accessible multipoles, would thus be highly valuable.
\section{SNR Calculations for Galaxy Surveys} 
\label{sec:formalism_snr}

In this section, we outline the formalism used to compute the signal-to-noise ratio (SNR) for individual bispectrum multipoles in the presence of massive neutrinos and also, explore the prospects for detection of these neutrino-induced signatures in the bispectrum for the \textit{Euclid} mission based on SNR. 

In the previous section, we quantified the modifications to the multipoles $Q_L^m$ in redshift-space induced by massive neutrinos. This involved computing the bispectrum multipoles using Eq.~\ref{eq:bispectrum_sh}, and incorporating corrections to the full redshift-space bispectrum as described in Eq.~\ref{eq:bispectrum_eq1}. Moreover, we have not included the neutrino corrections in the power spectrum term $P^r$, 
as our aim is to isolate and study their effects solely on the perturbation theory kernels $\tilde{F}$ and $\tilde{G}$.
However, in real galaxy surveys, what is measured is the total bispectrum, which inherently includes the effects of neutrinos in the power spectrum part. 
Therefore, in this analysis, we consistently include neutrino corrections at all the relevant places, ensuring compatibility with the real observations.
We use \texttt{CLASS} code and set $\sum m_\nu=0.12 \, {\rm eV}$ to generate neutrino corrected $P^r$ and incorporated this in the bispectrum along with the neutrino corrected PT kernels.
\subsection{Bispectrum Covariance}
\label{subsec:formalism}

Our results are presented in the context of the \textit{Euclid} galaxy survey \citep{Amendola:2016saw,Euclid:2019bue,Euclid:2019wjj,Euclid:2021qvm,Euclid:2019clj,Euclid:2021rez}. In any realistic survey, the number of available Fourier modes is finite due to the limited survey volume. In other words, the bispectrum can only be estimated from a discrete set of triangle configurations in Fourier space.
To account for this, we define the bispectrum multipole estimator as a discrete sum over all triangle configurations in Fourier space:

\begin{eqnarray}
\hat{B}_L^m (k_1,\mu, t) =   \sum_n \frac{w_L^m(\mathbf{\hat{p}}_n)}{2 V}  [\delta^s({\mathbf{k}}_n) \delta^s(\bar{\mathbf{k}}_n)
\delta^s(\tilde{\mathbf{k}}_n) +c.c.] \, ,
\label{eq:bispec_est}
\end{eqnarray}
where ``$c.c.$" represents complex conjugate and
$\mathbf{\hat{p}}_n$ quantifies the orientation of the $n^{\rm th}$ triangle w.r.t. the line-of-sight.
 $V$ is the survey volume and $n$ is the number of closed triangles. $w_L^m(\mathbf{\hat{p}}_n)$ is essentially a weight factor corresponding to the $n^{\rm th}$ triangle defined as,
\begin{eqnarray}
    w_L^m(\mathbf{\hat{p}}_n) = {\rm Re}\Big[\sqrt{2 L +1\over 4\pi}{ Y_L^m(\mathbf{\hat{p}}_n)
\over  \sum_{n_1} |Y_L^m(\mathbf{\hat{p}}_{n_1})|^2}\Big]\, .
\label{eq:weight}
\end{eqnarray}
Here the summation over $n$ takes care of all possible triangular configurations in a bin of ($k_1,\mu,t$) plane. The total number of $k$ modes within the interval $d^3k$ in a survey volume $V$ is simply given by, $dN_k=(2\pi)^{-3} \,V d^3k$. 
In ($k_1, \mu, t$) plane, the total number of triangles can be obtained as follows,
\begin{eqnarray}
    dN_{tr} = (2\pi)^{-6} \,V^2 d^3{k_1} d^3{k_2}\, .
\end{eqnarray}
All possible triangles within $d^3{k_1}$ can be obtained following rotations through Euler angles $\alpha$ and $\beta$, whereas total number of triangles within $d^3{k_2}$ can be obtained followed by a rotation about $\gamma$. Combining these, the total number of triangles can be computed giving,
\begin{eqnarray}
    dN_{tr} =(8\pi)^{-1} N_{tr} \, d\alpha \,\sin{\beta} \,d\beta \, d\gamma \, .
\end{eqnarray}
where $N_{tr}$ is the number of triangles within a bin of ($\Delta k_1, \Delta \mu, \Delta t$) centred around ($k_1,\mu,t$).
This can be given as follows,
\begin{eqnarray}
    N_{tr}=( 8 \pi^4)^{-1} \, \left(V k_1^3 \right)^2 \, t^2 \left[ \Delta  \ln k_1 \, ( t \, \Delta  \ln k_1  + \Delta t)  \, \Delta \mu \right] \, .
    \label{eq:no_triangle}
\end{eqnarray}
On the other hand, we have the closure relation that sets the normalization in  Eq.~\ref{eq:weight},
\begin{eqnarray}
 \sum_{n_1} \mid Y^m_{L} (\mathbf{\hat{p}}_{n_1}) \mid^2=(4 \pi)^{-1} N_{tr}\, .   
\end{eqnarray}
Further, in order to do a reasonable comparison with the CDM-only case,  we consider the following specifications: the bin width, $\Delta \ln k_1 =0.1, \Delta \mu =0.05$ and $\Delta t=0.05$ with $k_1$ is fixed to $0.2 \, {\rm Mpc}^{-1}$.
With this, the bispectrum estimator for different multipoles in Eq.~\ref{eq:bispec_est} leads to  the error covariance as,
\begin{eqnarray}
    C_{L L'}^{mm'} = \langle \Delta \hat{B}_L^m \, \Delta \hat{B}_{L'}^{m'}  \rangle \, ,
    \label{eq:cov}
\end{eqnarray}
where $\Delta \hat{B}_L^m $ is the statistical fluctuations defined as $\Delta \hat{B}_L^m = \hat{B}_L^m(k_1,\mu, t)  - \bar{B}^{m}_{L}(k_1,\mu,t)$. Here $\bar{B}^{m}_{L}(k_1,\mu,t)$ is the ensemble average of the estimator \textit{i.e.} $\bar{B}^{m}_{L}(k_1,\mu,t)=\langle \hat{B}_L^m(k_1,\mu, t) \rangle$. 
Finally given this, the error covariance for the bispectrum estimator can be calculated as,
\begin{eqnarray}
C_{L L'}^{mm'}(k_1,\mu,t) =\frac{ \sqrt{(2 L+1)(2 L^'+1)}}{N_{tr} }
\int d \Omega_{\mathbf{\hat{p}}}  {\rm Re}[Y_{L}^m(\mathbf{\hat{p}})] Re[Y_{L^'}^{m^'}  (\mathbf{\hat{p}})] \nonumber \\ \times \Big( 3 [B^s(k_1,\mu,t,\mathbf{\hat{p}})]^2  
+ V P^s(k_1,\mu_1) P^s(k_2,\mu_2) P^s(k_3,\mu_3) \Big)\,. 
\label{eq:cov_def2}
\end{eqnarray}

\subsubsection{Finger-of-God}
The error covariance estimated in the previous subsection does not take into account the FoG and Poisson noise. However,
at small scales, random motion of galaxies causes the density field to appear elongated along the line-of-sight (LoS), leading to a suppression of the observed clustering amplitude. This phenomenon is commonly referred to as the Finger-of-God (FoG) effect \citep{Jackson:1971sky}. Accounting for the FoG effect is essential for the scales probed in this analysis. The FoG effect is typically characterized by the pairwise velocity dispersion of galaxies. 
However, detailed analytical models to describe this are still lacking. 
Several phenomenological models have been proposed, typically introducing an ad-hoc 
damping factor based on the mean pairwise velocity dispersion. 
Among these, Lorentzian and Gaussian damping profiles are the most commonly used \citep{Jenkins:2000bv, Okumura:2015fga, Hikage:2015wfa}.
In this work, we adopt the Gaussian damping  to model the FoG-corrected power spectrum and bispectrum. Following \citet{Peebles:1980yev,Hikage:2015wfa,BaleatoLizancos:2025wdg},
the redshift-space power spectrum and bispectrum, including the FoG effect, can be related to their linear counterparts as follows,
\begin{eqnarray}\label{eq:pk_FoG}
P^s_{\rm FoG}(k_1,\mu_1) = 
 \exp{\left[(-k_1^2\mu_1^2){\sigma_p^2\over 2}\right]} P^s(k_1,\mu_1) 
\label{eq:pk_FoG}
\end{eqnarray}
\begin{eqnarray}
B^s_{\rm FoG}(\mathbf{k_1},\mathbf{k_2},\mathbf{k_3})= \exp \left[-(k_1^2\mu_1^2+k_2^2\mu_2^2+k_3^2\mu_3^2){\sigma_p^2\over 2}\right] \nonumber \\ \times B^s(\mathbf{k_1},\mathbf{k_2},\mathbf{k_3}) \, .
\label{eq:B_FoG}
\end{eqnarray}
where $P^s$ denotes the linear redshift-space galaxy power spectrum corrected for the Kaiser effect, and $B^s$ represents the corresponding induced bispectrum. The parameter $\sigma_p$ is the pairwise velocity dispersion of galaxies, expressed in comoving Mpc. In  general, $\sigma_p$ is treated as a free parameter. On very large scale where $k \sigma_p \ll 1$, Eqs.~\ref{eq:pk_FoG} and \ref{eq:B_FoG} reduce to the standard large-scale Kaiser limit \citet{Kaiser:1987qv}. In our analysis, we adopt the \textit{Euclid} galaxy survey specifications and use the corresponding value for the velocity dispersion \citep{Yankelevich:2018uaz}. 

\subsubsection{Shot Noise}
Galaxies are discrete tracers of the underlying matter density field. 
When discretizing this field, statistical fluctuations are introduced in the estimation of the N-point correlation functions—an effect known as shot noise. This phenomenon has been extensively studied in the context of large-scale structure \citep{Peebles:1980yev,Smith:2008ut,Matarrese:1997sk,Yankelevich:2018uaz,Yankelevich:2022mus,Ginzburg:2017mgf,Gualdi:2020ymf}. 
Shot noise corrections for the power spectrum, bispectrum, and higher-order N-point functions have been derived in \cite{Sugiyama:2019ike}.
Given the galaxy number density ($n_g$) in a survey volume $V$, the observed galaxy power spectrum and bispectrum can be expressed as,
\begin{eqnarray}
 V^{-1} P^s_{\rm obs}(k) & =& P^s(k,\mu_1) + P_{\rm shot}\, ,\\
\label{eq:pk_shot}
 V^{-1} B^s_{\rm obs}(\mathbf{k_1},\mathbf{k_2},\mathbf{k_3}) & =& B^s(\mathbf{k_1},\mathbf{k_2},\mathbf{k_3})  + {P^\prime}_{\rm shot} [P^s(k_1,\mu_1) \nonumber \\ &+ & P^s(k_2,\mu_2)+P^s(k_3,\mu_3)] + {{B^\prime}}_{\rm shot} \, ,
\label{eq:bk_shot}
\end{eqnarray}
where the shot noise contributions are given by: $P_{\rm shot} = {P^\prime}_{\rm shot}=n_g^{-1} $ and ${B^\prime}_{\rm shot}=n_g^{-2}$.
The corresponding change in the error covariance due to shot noise can be estimated as \cite{Mazumdar:2022ynd},
\begin{eqnarray} 
[C_{L L'}^{mm'}]_{\rm shot}-C_{L L'}^{mm'} \approx  V^{-1} [ n_g^{-5} +  n_g^{-4} P^s +  n_g^{-3} B^s ] \nonumber \\ + n_g^{-2} [P^s]^2  + n_g^{-1} P^s B^s \, ,
\label{eq:covg}
\end{eqnarray} 
where $[C_{L L'}^{mm'}]_{\rm shot}$ and $C_{L L'}^{mm'}$ are the covariance matrices with and without shot noise. 
We can estimate the contributions from different terms in Eq.~\ref{eq:covg} using realistic values for a \textit{Euclid}-like survey. At redshift $z = 0.7$, we assume: $n_g \sim 10^{-3}\, {\rm Mpc}^{-3}, \, P^s \sim 10^4 \, {\rm Mpc}^3, \, b_1 \sim 1, \, V \sim 10^9 \, {\rm Mpc}^9 $. The dominant contribution to the error covariance (excluding shot noise) comes from the second term of Eq.~\ref{eq:cov_def2} considering $k_1=0.2\,{\rm Mpc}^{-1}$, \textit{i.e.,}
$C_{L L'}^{m m'} \approx V [P^s]^3 \sim 10^{21} \, {\rm Mpc}^ {12} $. In contrast, the largest term in Eq.~\ref{eq:covg} involving shot noise is
$n_g^{-1} P^s B^s \sim 10^{15} \, {\rm Mpc}^{12}$, which is about five orders of magnitude smaller than the leading contribution to the error covariance. 
Therefore, the inclusion of shot noise has a negligible impact on the signal-to-noise (SNR) calculations for bispectrum multipoles within the $k_1$ range of our interest.
Although, in the presence of massive neutrinos, the shot noise terms can carry bispectrum multipole dependencies. Even so, their contribution remains subdominant. For completeness, we also present SNR forecasts including both FoG and shot noise effects in the following section.

\subsection{Signal-to-noise ratio}

Incorporating all these, the signal-to noise ratio (SNR) within the multipoles can be expressed in terms of the error covariance as,
\begin{eqnarray}
    {\rm SNR} = \frac{\left|\left({B_L^m|}_{f_\nu=0}-{B_L^m|}_{f_\nu \neq 0} \right)\right|}{\sqrt{{C_{LL}^{m m}|}_{f\nu \neq 0}}}\, .
    \label{eq:snr}
\end{eqnarray}
The signal in our analysis is defined as the difference between bispectrum multipoles in the presence and absence of massive neutrinos, while the covariance matrix is computed following Eq.~\ref{eq:cov_def2}, including the effects of massive neutrinos. 
The FoG effect is incorporated in the covariance matrix via Eqs.~\ref{eq:pk_FoG} and \ref{eq:B_FoG}, while shot noise is included using Eq.~\ref{eq:covg}.

In what follows, we discuss the signal-to-noise predictions for the multipoles in presence of massive neutrinos. 
%
\section{Detection Prospects with \textit{Euclid}}
\label{subsec:snr_prediction}

Our focus in this section is to forecast the signal-to-noise ratio (SNR) for bispectrum multipoles in the context of the \textit{Euclid} galaxy survey \citep{2010arXiv1001.0061R,Amendola:2016saw,Euclid:2024yvv,Euclid:2024sqd,Euclid:2024yrr} using the above-mentioned analysis. We consider the signal to be statistically significant whenever ${\rm SNR} \geq 1$. 
For the present analysis, we adopt conservative values for survey specifications and bias parameters at redshift $z=0.7$ \citep{Yankelevich:2018uaz}. Specifically, we use: $b_1 =1.18, \, b_2 = -0.76, \, V = 9.04\, {\rm Gpc}^3$ and $\sigma_p = 7.09 \, {\rm Mpc}, \, n_g = 8.6 \times 10^{-4}$.
Note that the matter power spectrum is extracted from \texttt{CLASS} \citep{2011JCAP...07..034B}, employing the same cosmological parameters described earlier in Sec.~\ref{subsec:real_space_bispectrum_with_nu} along with $\sum m_\nu=0.12 \, {\rm eV}$ . From this point onward, in addition to the modifications of the PT kernels, we incorporate all contributions from neutrinos. For the analysis, we adopt bin widths of $\Delta \ln k_1 =0.1, \, \Delta \mu =0.05, \, \Delta t =0.05$. We will later assess how this choice of binning impacts the multipoles of interest. Throughout the analysis, we primarily focus on statistically significant multipoles.

\begin{figure*}
    \centering
    \subfloat{\includegraphics[width=0.55\columnwidth]{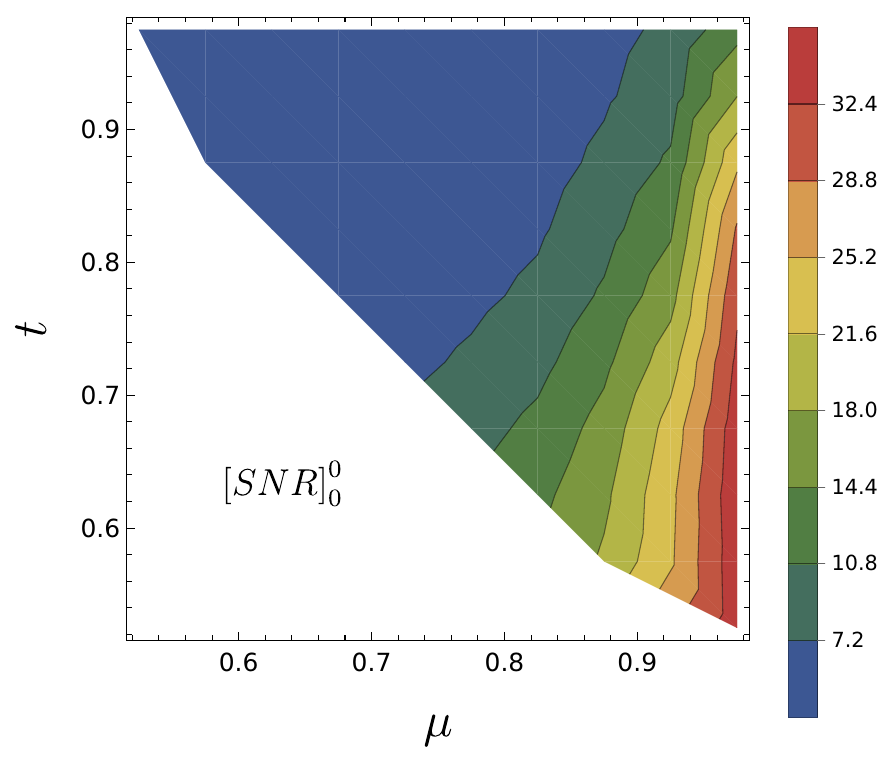}}
    \subfloat{\includegraphics[width=0.55\columnwidth]{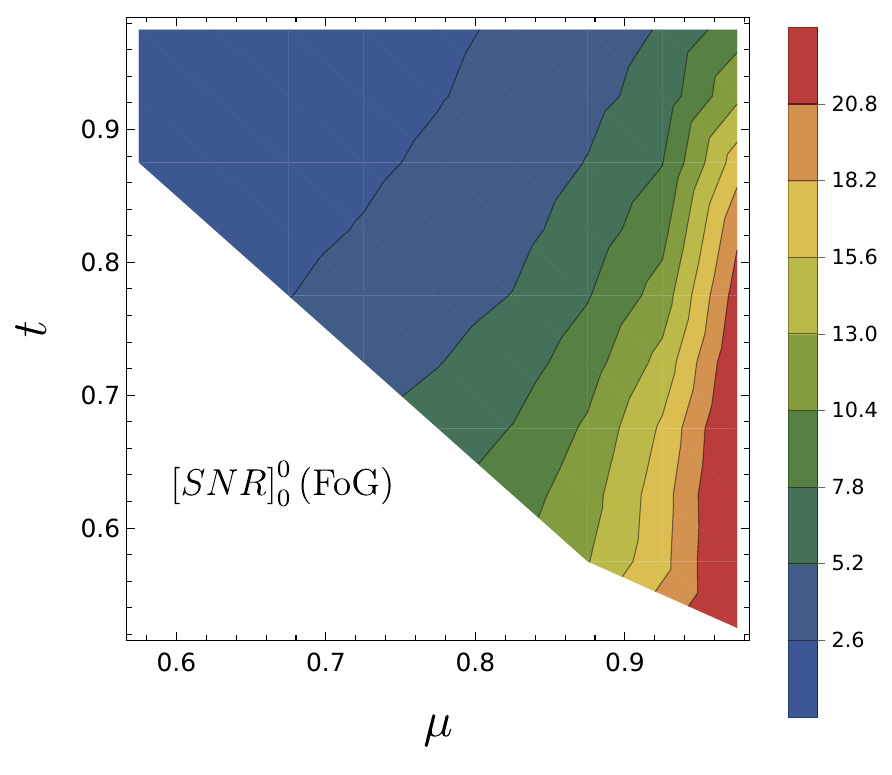}}
    \subfloat{\includegraphics[width=0.55\columnwidth]{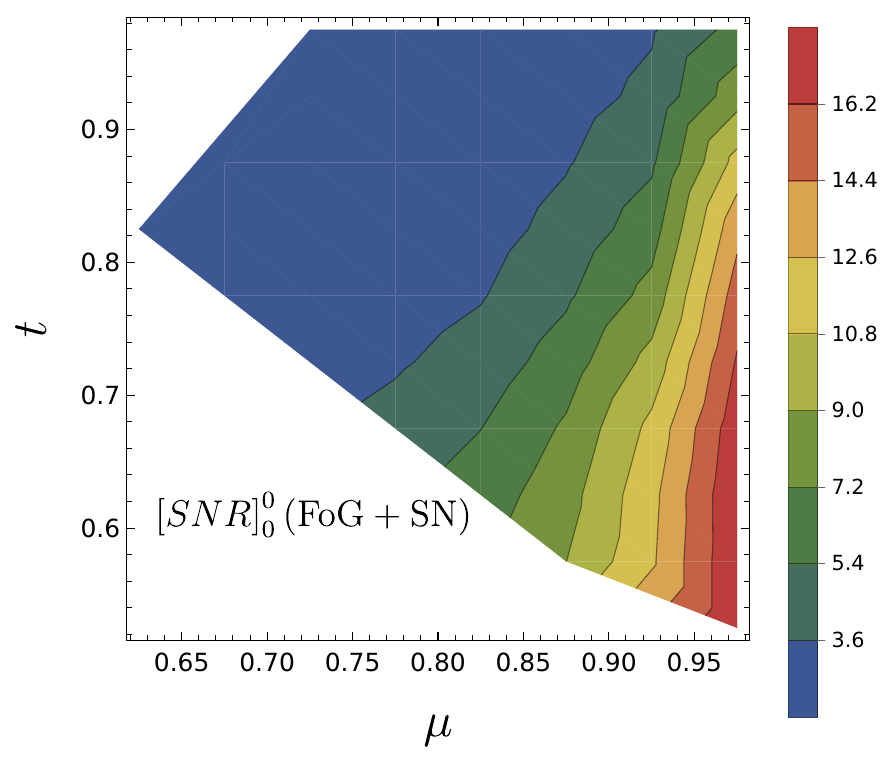}}
    \quad
    \subfloat{\includegraphics[width=0.55\columnwidth]{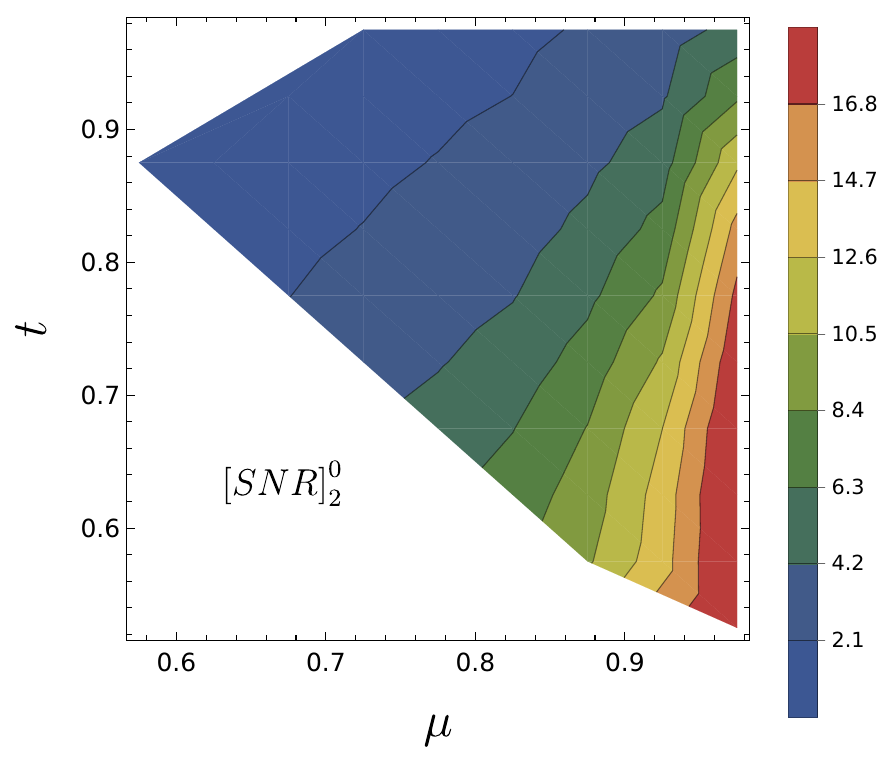}}
    \subfloat{\includegraphics[width=0.55\columnwidth]{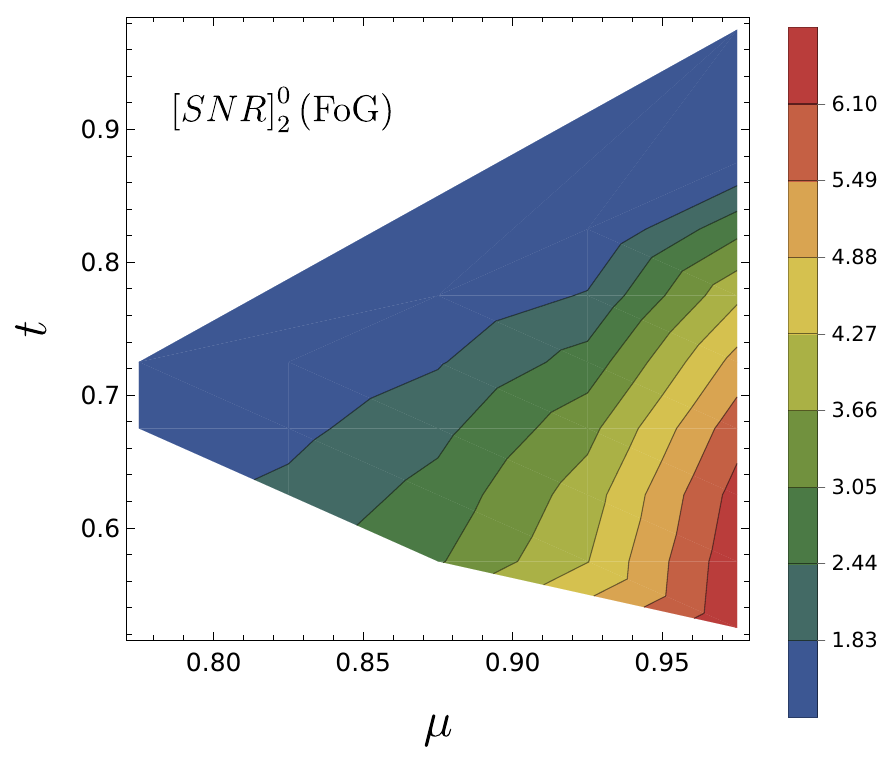}}
    \subfloat{\includegraphics[width=0.55\columnwidth]{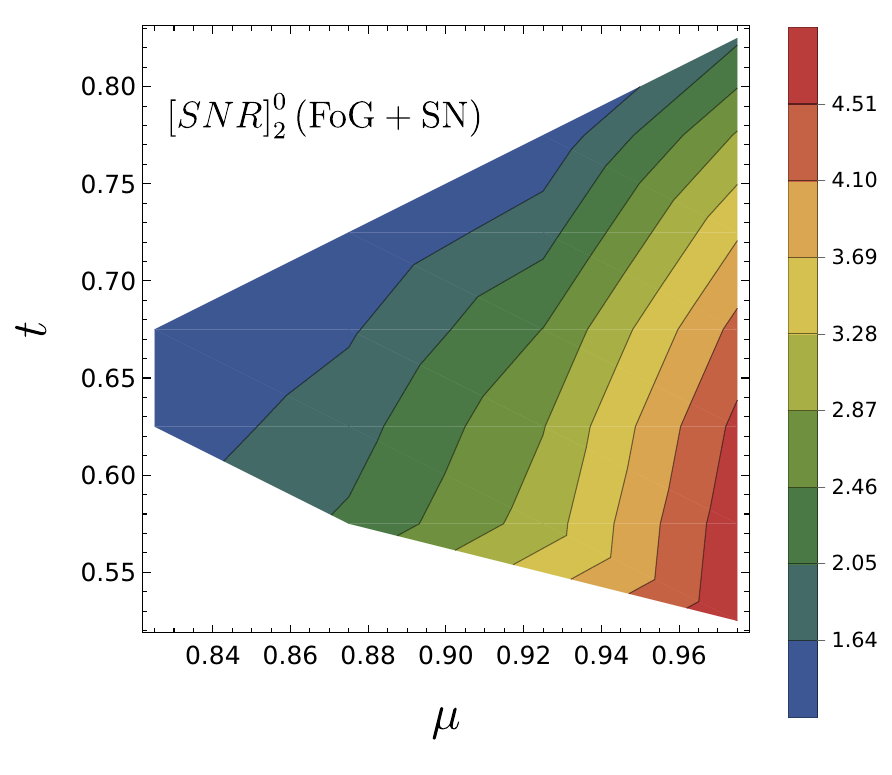}}
    \caption{Signal-to-noise ratio (SNR) maps for bispectrum multipoles $B_0^0$ and $B_2^0$ at $k_1 = 0.2\,\mathrm{Mpc}^{-1}$ and $z = 0.7$, shown under three scenarios: no FoG damping or shot noise (\textit{left}), FoG only (\textit{middle}), and FoG + shot noise (\textit{right}). Colorbars indicate SNR $\geq 1$. Both multipoles are robustly detectable in linear triangle configurations, but shot noise significantly reduces the SNR, particularly for squeezed shapes.
    }
    \label{fig:snr_1}
\end{figure*}

The top and bottom panels of Fig.~\ref{fig:snr_1} present the SNR predictions for $B_0^0$
and $B_2^0$, respectively. In each case, the left panels show predictions without the FoG effect, the middle panels include FoG, and the right panels incorporate both FoG and shot noise. Across all panels, we observe that the SNR is highest for linear triangles ($\mu \approx 1$), increasing further as we approach the limit ($\mu \rightarrow 1, t \rightarrow 0.5$). 
For both multipoles, the SNR is maximized in the left panels, where neither FoG damping nor shot noise is included. In particular, the maximum SNR in $B_0^0$ and $B_2^0$ without incorporating any effects is approximately $34$ and $20$ at ($\mu \to 1,t \approx 0.525$).
Introducing the FoG effect leads to a significant reduction in SNR - by approximately a factor of 1.6 for $B_0^0$
and a factor of 2.8 for $B_2^0$. The SNR decreases further, though more moderately (1.9 times and 4 times), when both FoG and shot noise are considered. 
These results are consistent with previous studies, where only the monopole is analyzed \citep{Hahn:2020lou}.
Overall, for both multipoles, the SNR for equilateral triangles remains \( \lesssim 1 \), suggesting that these configurations are not optimal for probing neutrino mass. However, we find a 
large number of triangles with $\mu>t$ to have ${\rm SNR}>5$ and can be crucial for detecting
massive neutrinos.

\begin{figure*}
    \centering    
    \subfloat{\includegraphics[width=0.55\columnwidth]{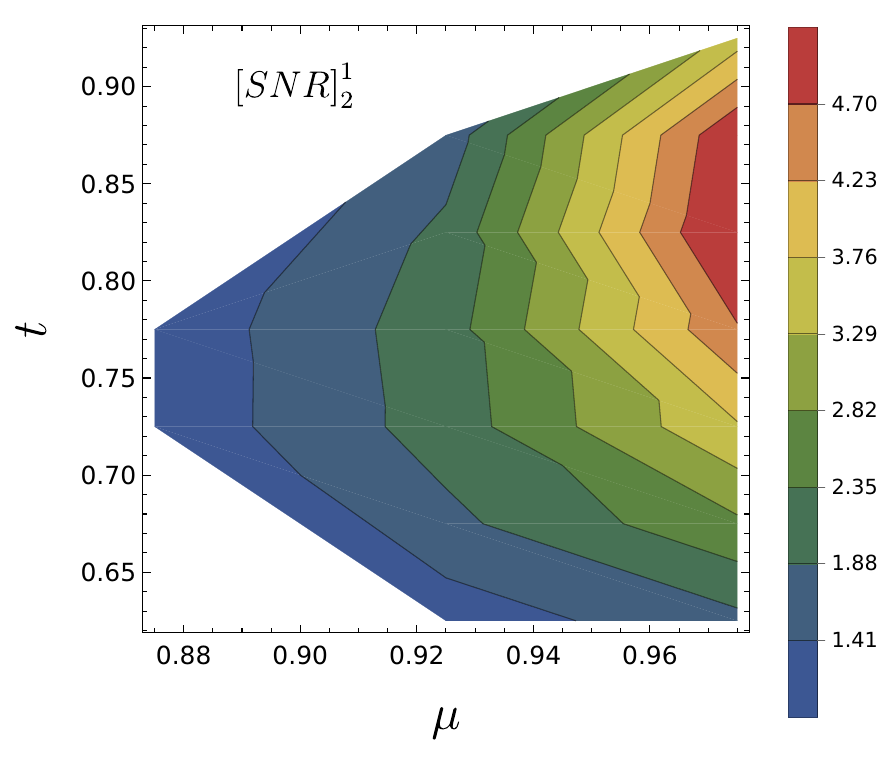}}
    \subfloat{\includegraphics[width=0.55\columnwidth]{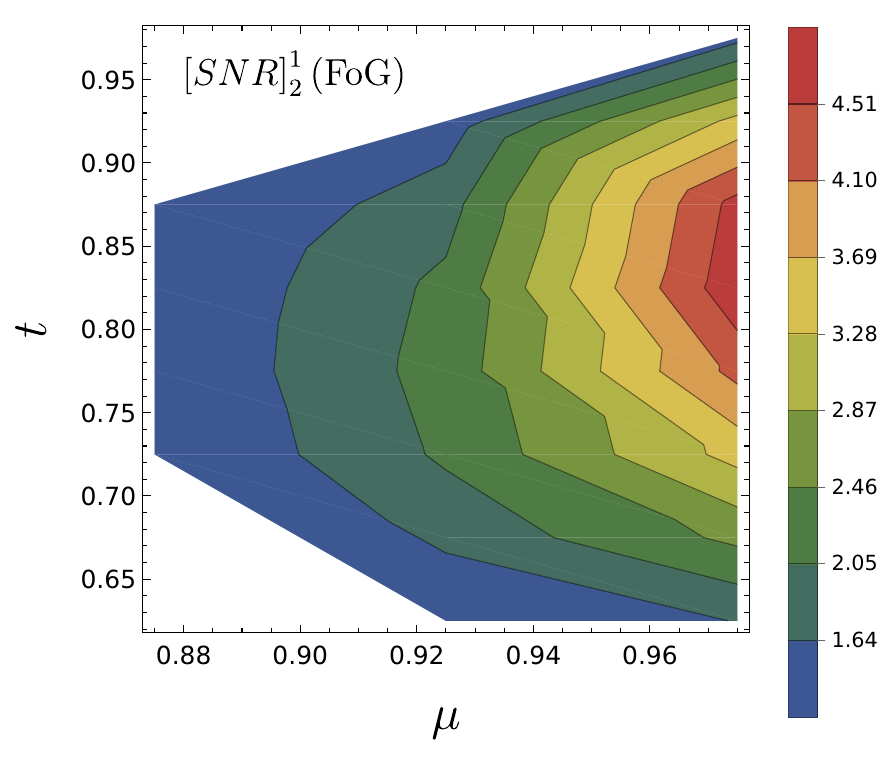}}
    \subfloat{\includegraphics[width=0.55\columnwidth]{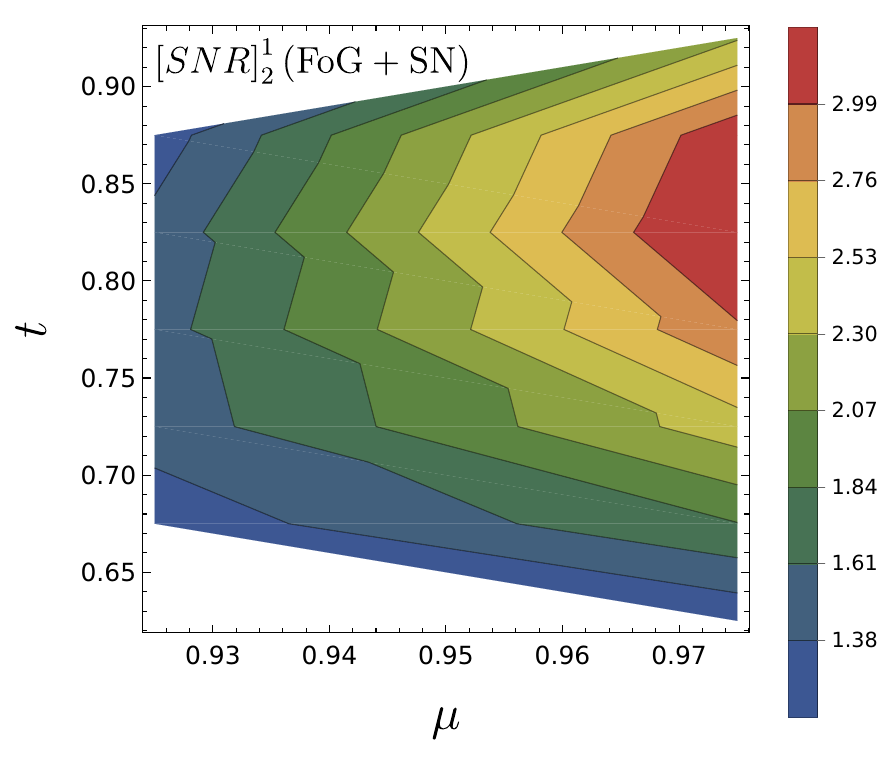}}
    \quad
    \subfloat{\includegraphics[width=0.55\columnwidth]{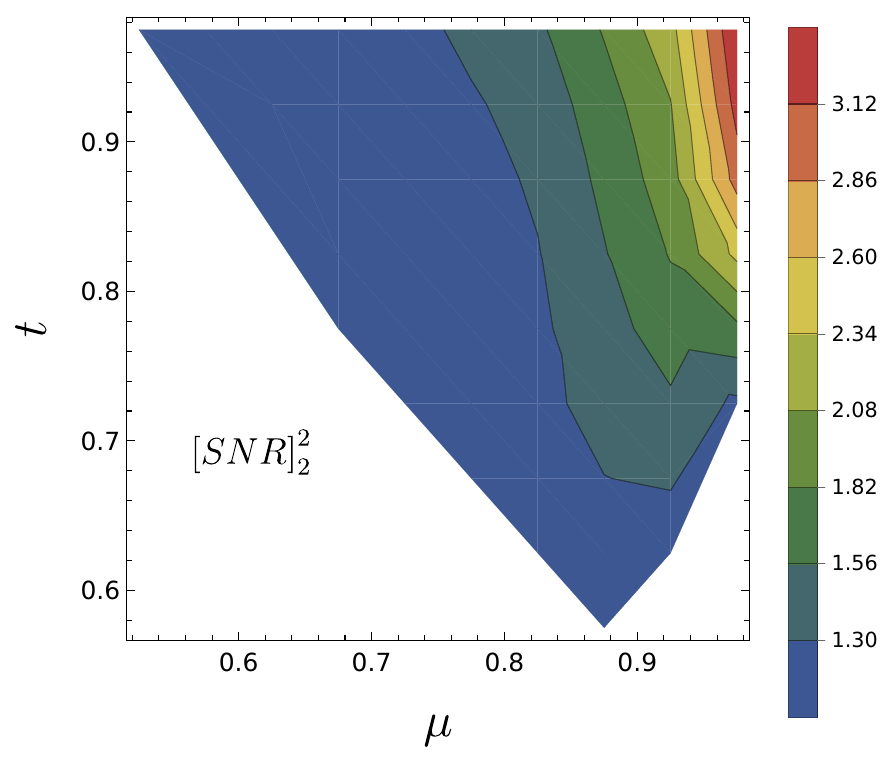}}
    \subfloat{\includegraphics[width=0.55\columnwidth]{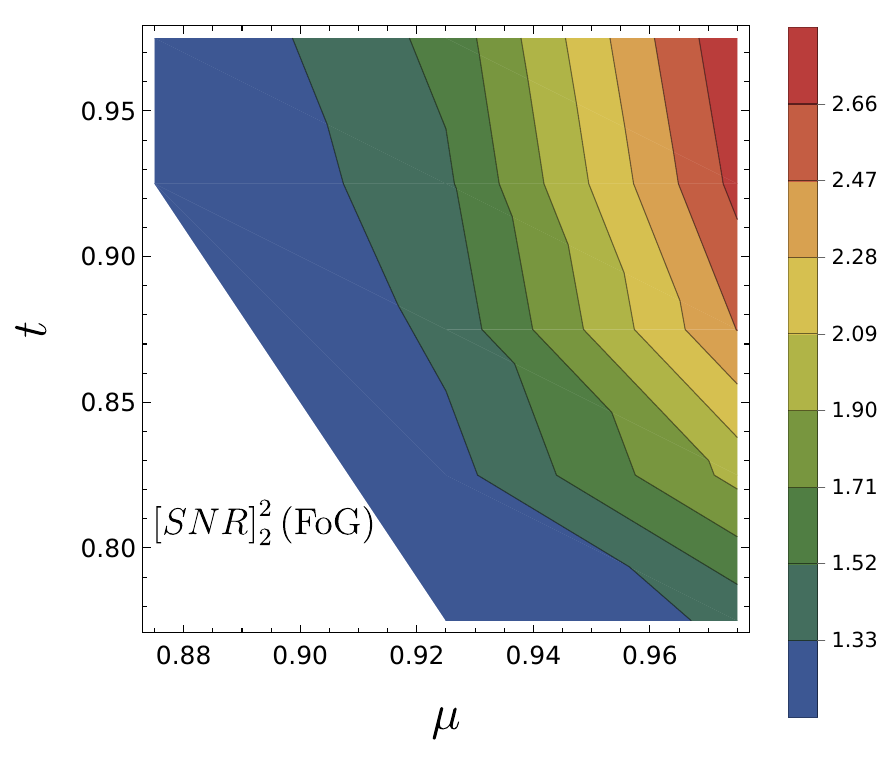}}
    \subfloat{\includegraphics[width=0.55\columnwidth]{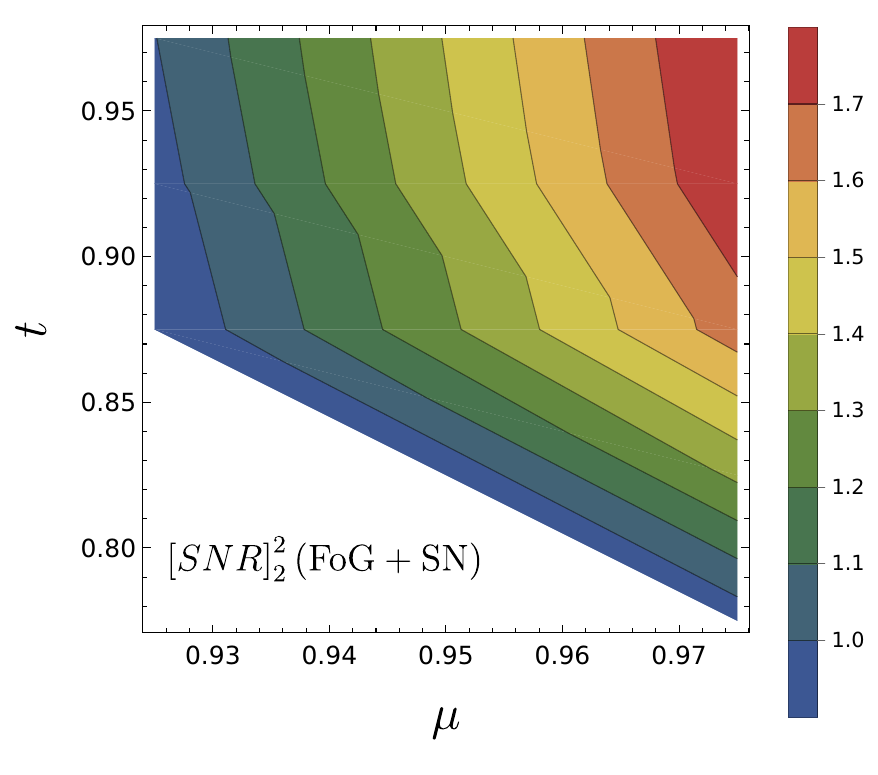}}
    \caption{SNR maps for higher multipoles $B_2^1$ and $B_2^2$ under the same setups as 
    Fig.~\ref{fig:snr_1}. Although the overall SNR values are lower compared to $B_0^0$ and $B_2^0$, significant detections ($\mathrm{SNR} > 2$) persist for squeezed and stretched triangle configurations. Detection prospects for $B_2^1$ are stronger than those for $B_2^2$.}
    \label{fig:snr_2}
\end{figure*}

The top and bottom panels of Fig.~\ref{fig:snr_2} show the SNR predictions for $B^1_2$ and $B^2_2$, respectively. As in Fig.~\ref{fig:snr_1}, the left panels present results without the FoG effect, the middle panels include FoG, and the right panels account for both FoG and shot noise. We observe that the SNR is again highest for linear triangle configurations ($\mu \approx 1$), where the SNR for 
$B^1_2$ peaks around $t\approx0.8$, and for $B^2_2$ it peaks near the squeezed limit $t\approx1$.
The maximum SNR ($\approx 7 $ for both $B^1_2$ and $B^2_2$ ) is achieved in the absence of FoG and shot noise, shown in the left panels. 
Including FoG introduces a very small reduction in SNR, by a factor of $\approx1.04$, 
for $B^1_2$, while the reduction is moderate, by a factor of $\approx1.5$ for $B^2_2$.
The addition of shot noise causes a further reduction in SNR, by a 
factor of $\approx1.5$, for both the multipoles. Compared to Fig.~\ref{fig:snr_1}, the overall SNR values for $B^1_2$ and $B^2_2$ are lower, indicating that these multipoles are less sensitive than $B_0^0$ and $B_2^0$ for neutrino mass detection. 
However, considering $B^1_2$, a lot of triangles can still be found with ${\rm SNR}>3$, which 
may provide significant information regarding the neutrino mass. Considering $B^2_2$, 
a number of triangles near the squeezed limit exhibit ${\rm SNR}>2$, exhibiting their 
significance in neutrino mass detection.
These results highlight that while $L=2$ multipoles can contribute additional information, their constraining power on neutrino mass is more limited compared to $L=0$ multipoles.
\begin{figure*}
    \centering   
    \subfloat{\includegraphics[width=0.55\columnwidth]{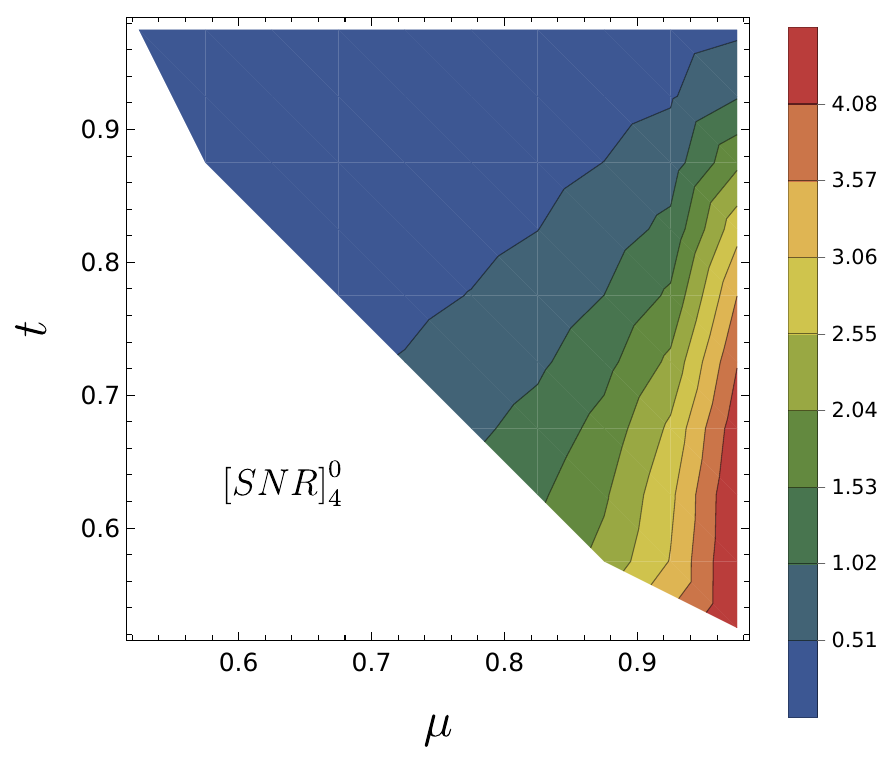}}
    \subfloat{\includegraphics[width=0.55\columnwidth]{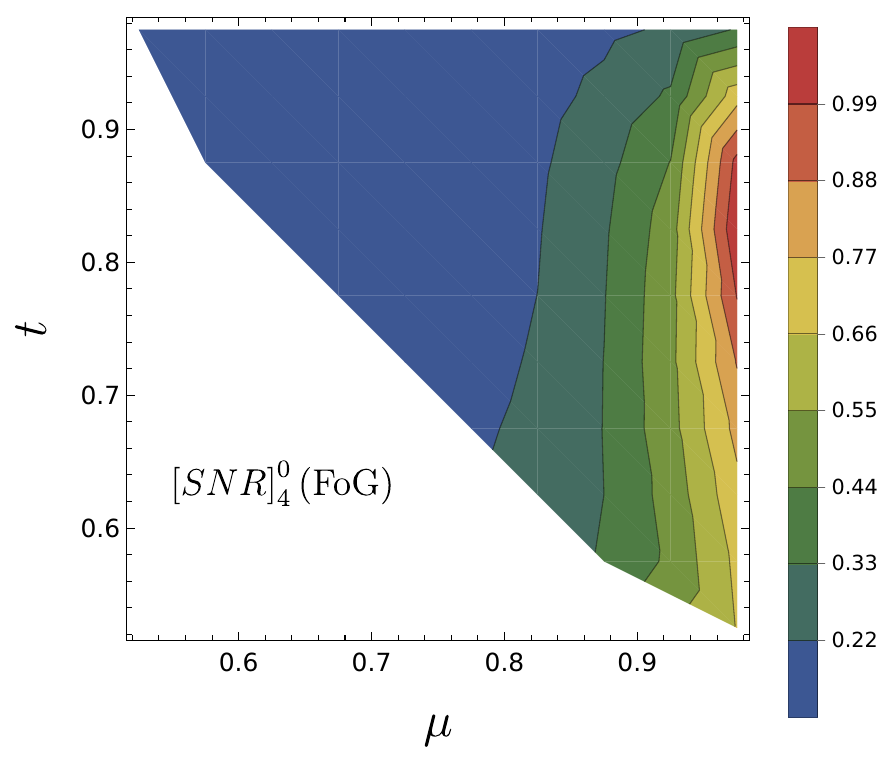}}
    \subfloat{\includegraphics[width=0.55\columnwidth]{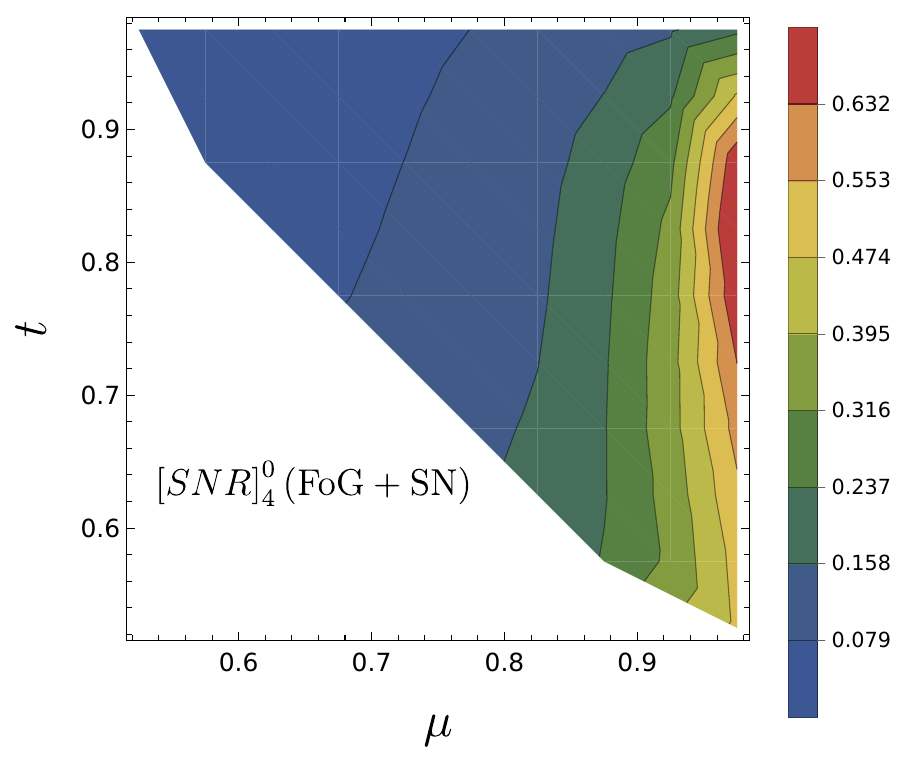}}
    \quad
    \subfloat{\includegraphics[width=0.55\columnwidth]{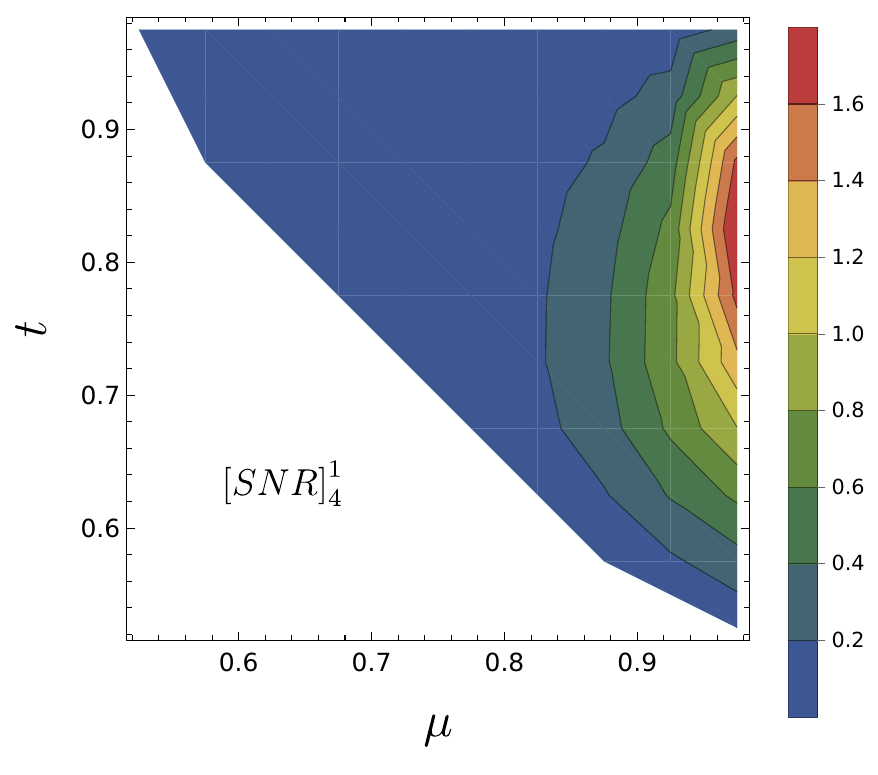}}
    \subfloat{\includegraphics[width=0.55\columnwidth]{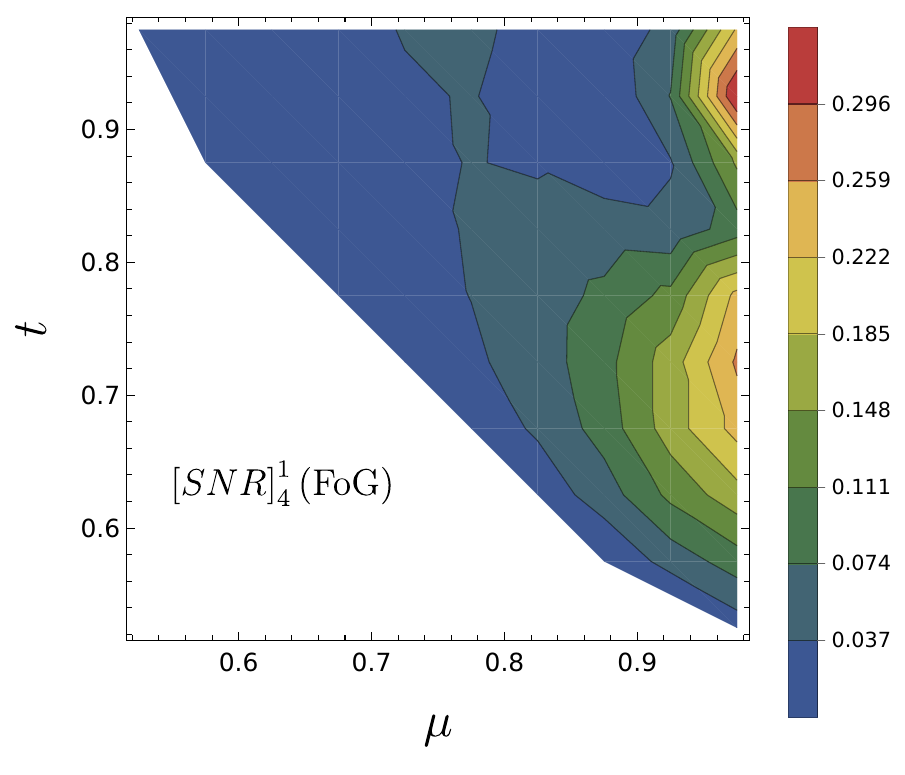}}
    \subfloat{\includegraphics[width=0.55\columnwidth]{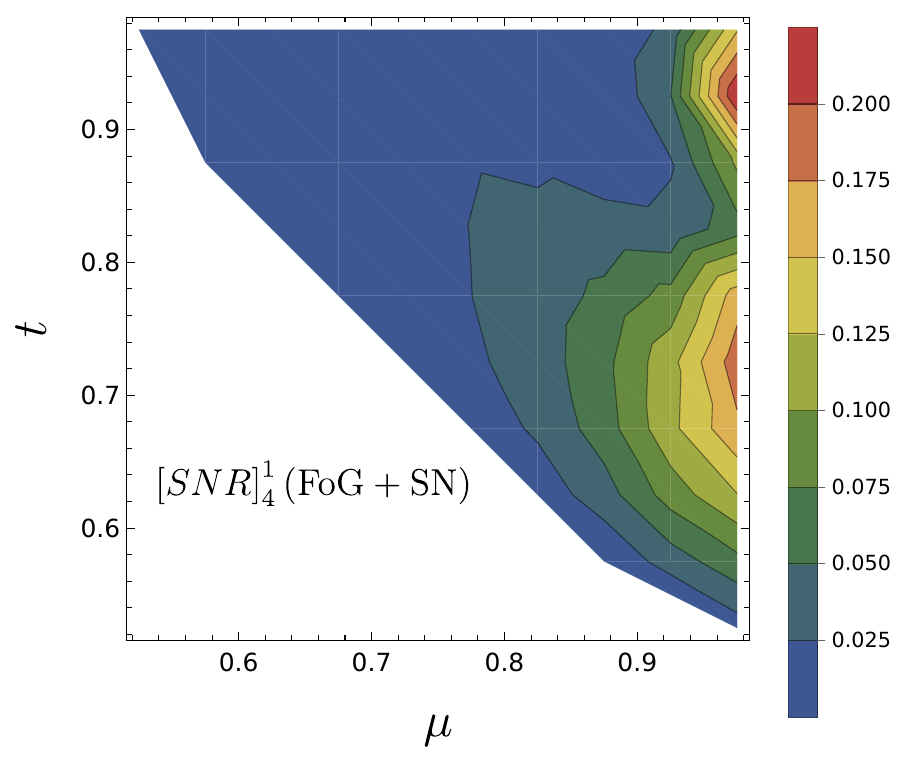}}
    \caption{SNR maps for higher-order multipoles $B_4^0$ and $B_4^1$ at $k_1 = 0.2\,\mathrm{Mpc}^{-1}$ and $z = 0.7$, shown with: no FoG damping (\textit{left}), FoG only (\textit{middle}), and both FoG and shot noise (\textit{right}). Without shot noise, $\mathrm{SNR} > 1$ is achieved for linear triangle configurations. However, with realistic shot noise and FoG damping, the SNR drops below unity across most of the $(\mu, t)$ plane, indicating difficulty in detecting higher-order multipoles.
    }
    \label{fig:snr_3}
\end{figure*}

\begin{figure*}
    \centering
    \subfloat{\includegraphics[width=0.55\columnwidth]{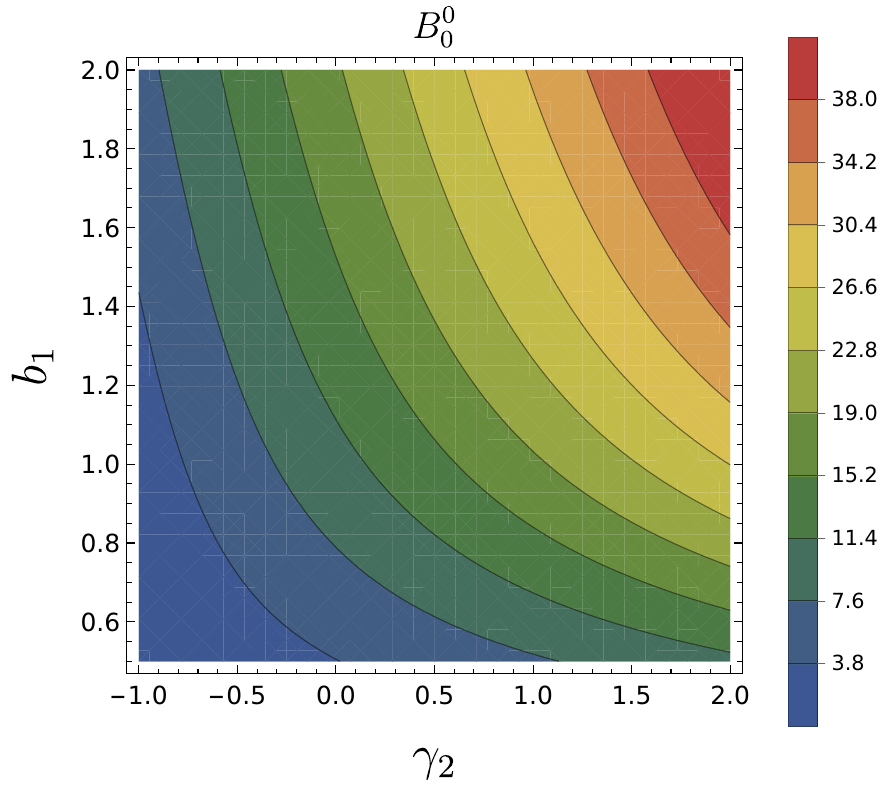}}
    \subfloat{\includegraphics[width=0.55\columnwidth]{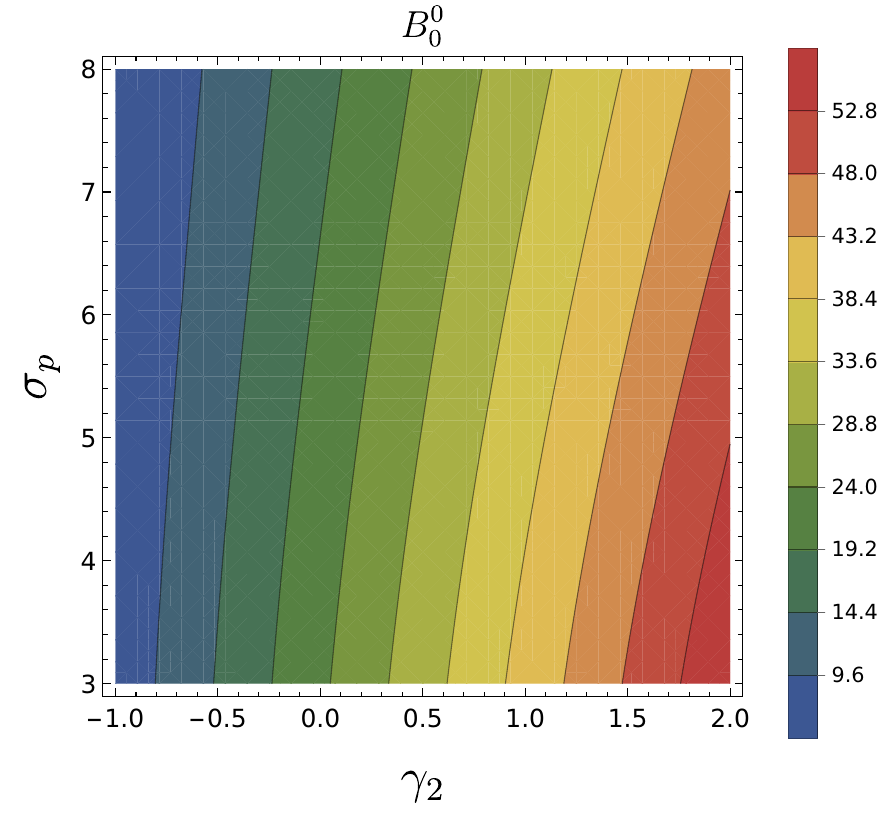}}
    \subfloat{\includegraphics[width=0.55\columnwidth]{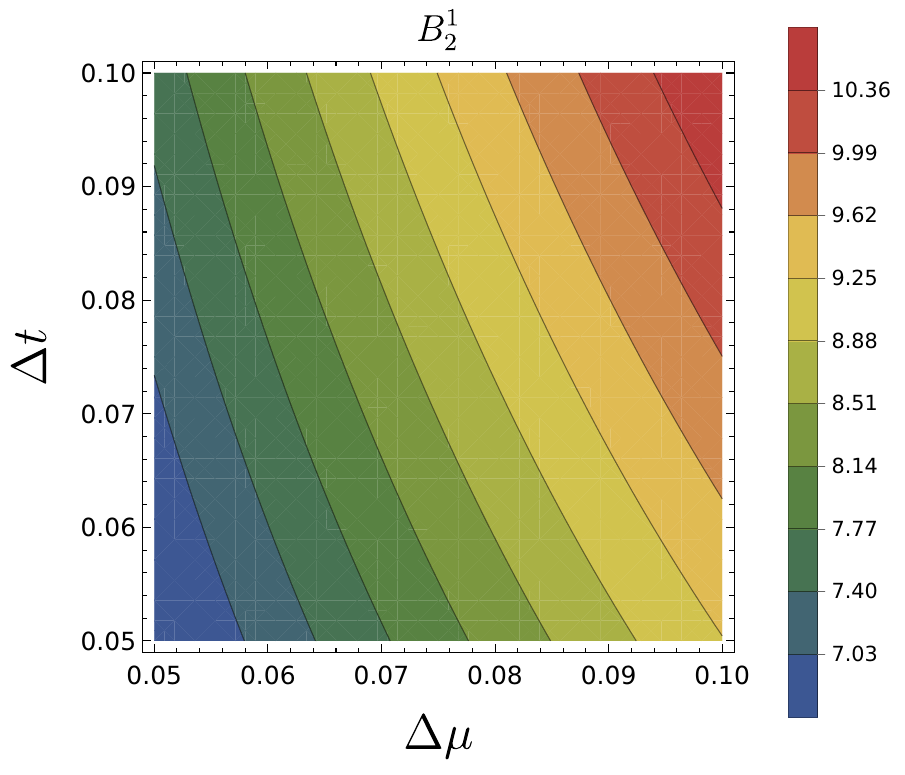}}
    \quad
    \subfloat{\includegraphics[width=0.55\columnwidth]{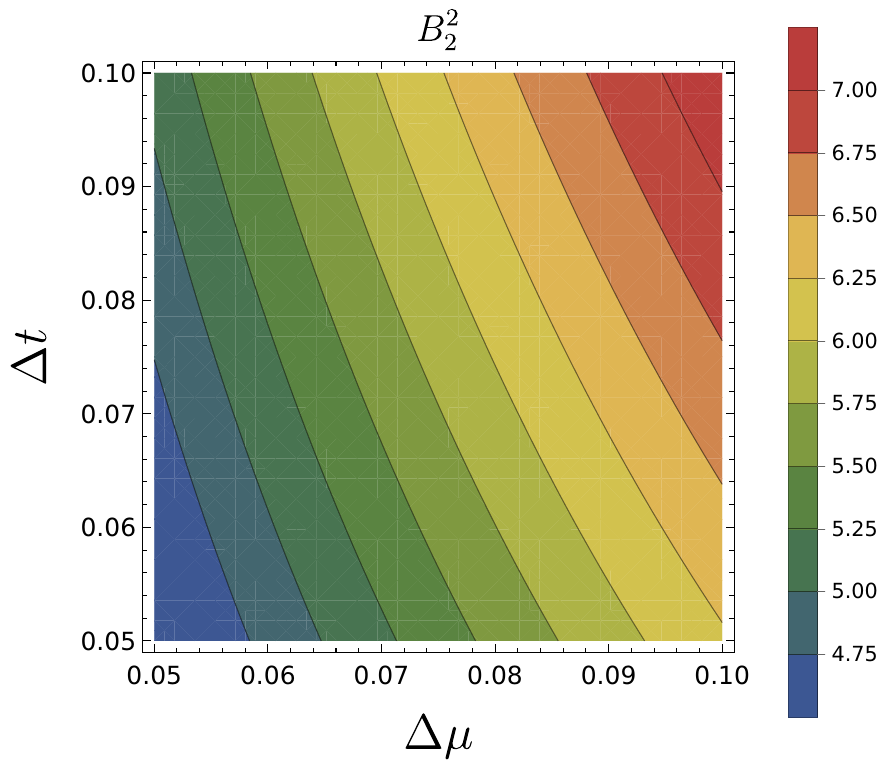}}
    \subfloat{\includegraphics[width=0.55\columnwidth]{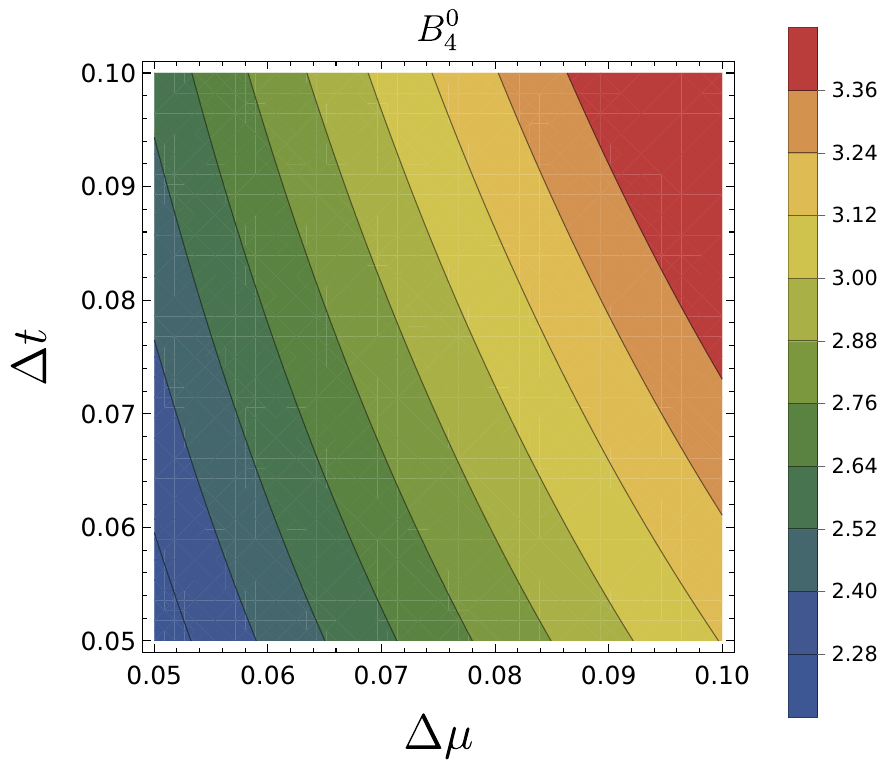}}
    \subfloat{\includegraphics[width=0.55\columnwidth]{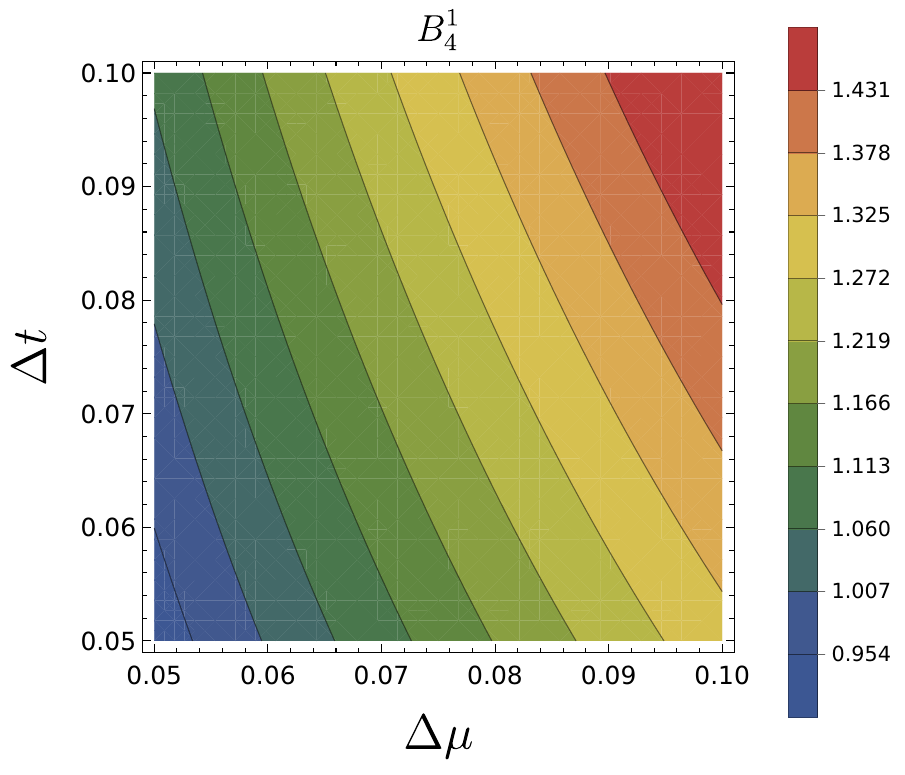}}
    \caption{In the \textit{upper left } and \textit{central panel} we present the dependency of SNR on $\gamma_2- b_1$ and $\gamma_2-\sigma_p$ pairs 
    for  $B_0^0$ multipole. This results are shown for \texttt{squeezed} triangle shape ($\mu=0.975$, $t=0.975$) at fixed $k_1=0.2 \, {\rm Mpc}^{-1}$ and redshift $z=0.7$. In the \textit{upper right panel}, we show the binning dependence of SNR for $B_2^1$ multipole. The same for $B_2^2$ multipole is shown in the \textit{lower left panel} at ($\mu \to 1, t =0.975$). In \textit{lower central} and \textit{right panel}, binning dependency of SNR for $B_4^0$ and $B_4^1$ is shown for squeezed triangles with $\mu \to 1$, $t \approx 0.975$. All these plots are generated with fixed  $k_1=0.2 \, {\rm Mpc}^{-1}$ and redshift $z=0.7$.}
    \label{fig:case_study}
\end{figure*}

\begin{figure*}
    \centering
    \subfloat{\includegraphics[width= 1.8 \columnwidth]{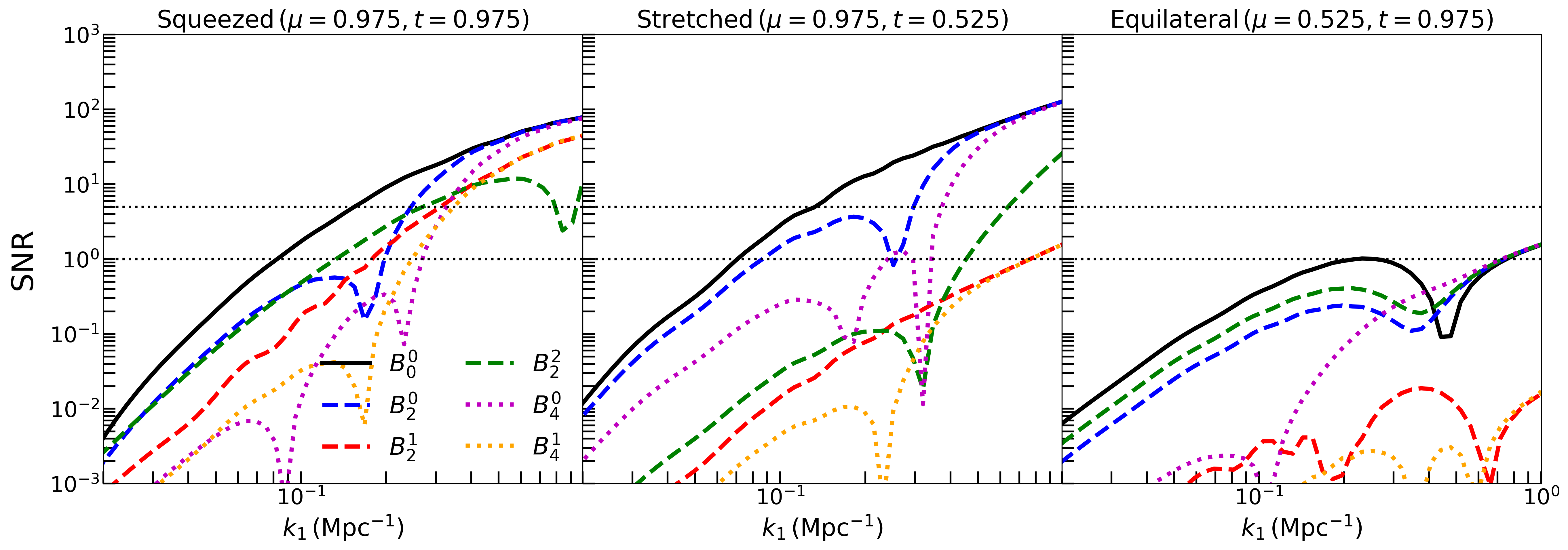}}
    \caption{$k_1$ dependence of SNR for $B_0^0$, $B_2^0$, $B_2^1$, $B_2^2$, $B_4^0$, and $B_4^1$ multipoles at fixed redshift $z=0.7$. We present results for specific triangle shapes: \texttt{squeezed} (left panel), \texttt{stretched} (central panel) and \texttt{equilateral} (right panel).}
    \label{fig:snr_k1}
\end{figure*}

\begin{figure*}
    \centering
    \subfloat{\includegraphics[width=0.9 \columnwidth]{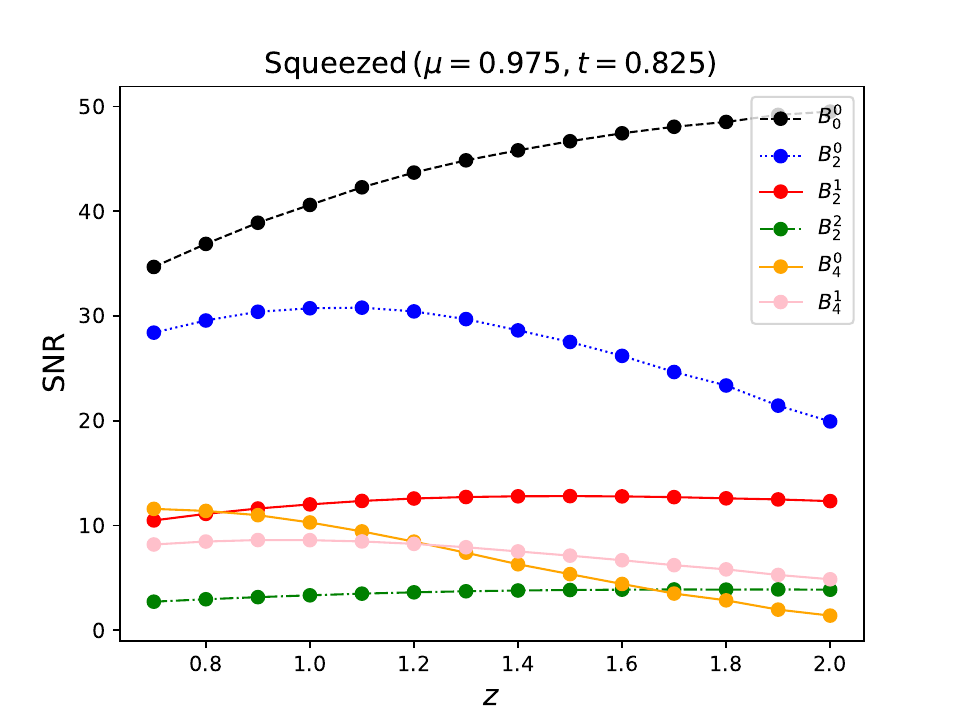}}
    \caption{ Redshift dependence of SNR for 
     $B_0^0$, $B_2^0$, $B_2^1$, $B_2^2$, $B_4^0$, and $B_4^1$ multipoles. We show results for \texttt{squeezed} triangle ($\mu=0.975, t =825$) at a fixed $k_1 = 0.4\, {\rm Mpc}^{-1}$. Filled circles represent redshift ($z$) bins adopted from \citep{Yankelevich:2018uaz}.}
    \label{fig:snr_z}
\end{figure*}

The top and bottom panels of Fig.~\ref{fig:snr_3} show the SNR predictions for the multipoles 
$B^0_4$ and $B^1_4$, respectively. For both, the SNR is maximum along the linear triangles in
all scenarios. Considering $B^0_4$, the SNR is maximum without FoG and shot noise and peaks 
at ($\mu\approx1, t\approx0.5$) where the value is $\approx5$. When FoG and shot noise are 
introduced, the overall SNR reduces significantly. The maximum is shifted close to 
($\mu\approx1, t\approx0.8$) and the value goes below unity.
The same is true for $B^1_4$, only the maximum occurs at different locations. Also, the SNR, even without FoG and shot noise, is $<2$ for this multipole, and this goes well below unity when FoG and shot noise are considered. These results suggest that the hexadecapole moments carry relatively little constraining power for neutrino mass detection compared to the lower-order multipoles.

In the top two panels of Fig.~\ref{fig:case_study}, we study how the SNR for the monopole $B_0^0$ 
in the squeezed limit ($\mu\rightarrow1, t=0.975$)
varies with different physical parameters: the linear bias $b_1$, the quadratic bias quantified by $\gamma_2$, and the pairwise velocity dispersion $\sigma_p$. 
In all previous figures, we adopted a single fiducial value for each of these parameters. However, since the SNR depends sensitively on these quantities, we now explore how changes in them impact the results. From the left panel, we observe that the SNR increases steadily with increasing $b_1$ and $\gamma_2$, indicating that a larger bias or positive non-linear corrections enhance the signal. For negative $\gamma_2$, the increase in SNR with $b_1$ is slower compared to positive $\gamma_2$ values. Similarly, for small $b_1$, the growth of SNR with $\gamma_2$ is less pronounced than for larger $b_1$. In the middle panel, we find that the SNR increases with $\gamma_2$ but decreases with $\sigma_p$, although the dependence on $\sigma_p$ is somewhat milder. 
A higher $\sigma_p$ dampens the clustering amplitude through the FoG effect, and this 
reduces the SNR.
For a fixed $\gamma_2$, the rate at which SNR decreases with $\sigma_p$ becomes steeper as $\gamma_2$ increases. 
These trends are broadly consistent for other multipoles as well, although the overall SNR values vary depending on the specific multipole considered. These results suggest that variations in the bias parameters and the velocity dispersion can significantly impact the detectability of neutrino mass. 

The top right and bottom panels of Fig.~\ref{fig:case_study} show the impact of binning choices on the SNR for the multipoles $B_2^1$, $B_2^2$, $B_4^0$, and $B_4^1$ in the squeezed limit 
($\mu\rightarrow1, t=0.975$). 
In these panels, we vary the bin widths $\Delta t$ and $\Delta \mu$ and plot the resulting 
SNR for each multipole. In our previous setup, higher-order multipoles exhibit 
significantly lower SNR, motivating us to investigate whether optimizing the binning 
could enhance the effective SNR and thus allow more information about neutrino mass to be extracted from these components. Across all four panels, we find that the SNR systematically increases 
as the bin widths $\Delta t$ and $\Delta \mu$ are enlarged. This behavior is expected: 
wider bins group together more triangle configurations, effectively summing their signals and increasing the total SNR, even though finer details in the shape dependence are lost. 
In particular, we observe that increasing $\Delta \mu$ has a stronger impact on the 
SNR than increasing $\Delta t$. This suggests that the SNR in these multipoles is 
especially sensitive to the binning along the $\mu$ direction. 
The improvement is especially important for $B_4^0$ and $B_4^1$, where the initial 
SNR was very low (close to or below unity for narrow bins). By adopting moderate bin sizes (e.g., $\Delta \mu, \Delta t \sim 0.1$), the SNR for these higher-order multipoles can be increased 
by factors of $2\text{--}3$, partially recovering the lost statistical power. However, this comes with a 
trade-off: while coarser binning boosts detectability, it reduces our ability to resolve detailed 
shape dependencies, which could affect the interpretation of neutrino mass constraints if the 
signal varies strongly with triangle configuration.
Overall, these results suggest that careful optimization of binning strategies is crucial when attempting to include higher multipole moments in cosmological analyses targeting neutrino mass.

Fig.~\ref{fig:snr_k1} shows the variation of SNR as a function of $k_1$ at a fixed redshift 
$z = 0.7$ for three different triangle configurations: squeezed, stretched, and equilateral. 
The SNR is plotted for six different multipoles: $B_0^0$, $B_2^0$, $B_2^1$, $B_2^2$, $B_4^0$,
and $B_4^1$. The horizontal dotted lines indicate the $1\sigma$ and $5\sigma$ significance thresholds. All results include the effects of Finger-of-God (FoG) damping.

In the squeezed triangle configuration ($\mu = 0.975, t = 0.975$), the SNR increases steadily with increasing $k_1$ for all multipoles. The $B_0^0$ multipole reaches the highest SNR values across the $k_1$ range. Some multipoles, especially $B_2^0$, $B_4^0$ and $B_4^1$, show visible dips around 
$k_1 \sim 0.2\text{--}0.3\,{\rm Mpc}^{-1}$, associated with zero crossings in the bispectrum signal. 
Although small at small $k_1$, 
$B_4^0$ and $B_4^1$ achieve large SNR values, around 60 and 35, respectively
at $k_1 \sim 0.8\,{\rm Mpc}^{-1}$. The squeezed configuration is particularly promising for probing neutrino mass signatures at $k_1 > 0.1\,{\rm Mpc}^{-1}$ in almost all the multipoles.
In the stretched triangle configuration ($\mu = 0.975, t = 0.525$), a similar trend of increasing SNR with $k_1$ is observed. SNR is higher for the $B_0^0$, $B_2^0$, and $B_4^0$ multipoles. Sharp dips around $k_1 \sim 0.3\text{--}0.4\,{\rm Mpc}^{-1}$ are again visible. 
Considering $L=2$, $B_2^2$ has significantly more SNR compared to $B_2^1$ at $k_1 > 0.3\,{\rm Mpc}^{-1}$.
In the equilateral triangle configuration ($\mu = 0.525, t = 0.975$), the SNR values are generally lower across all multipoles compared to the other two configurations. Although the SNR increases with $k_1$, only $B_0^0$ crosses the $5\sigma$ threshold at high $k_1$, while the other multipoles remain significantly below this level. The difference between the neutrino-corrected and standard bispectrum appears too small to achieve detection in equilateral configurations, and increasing the bin width does not substantially enhance the SNR.

Overall, we find that SNR improves with $k_1$ across all triangle configurations, but the growth rate and dominant multipoles depend sensitively on the triangle shape. Squeezed triangles are most promising for detecting neutrino signatures, followed by the stretched triangles. The behavior of SNR with $k_1$ plays a critical role in recovering neutrino information, especially from higher-order multipoles.

In Fig.~\ref{fig:snr_z}, we show the dependency of SNR on redshift for different multipoles, 
considering a squeezed triangle configuration with $\mu = 0.975$ and $t = 0.825$ at fixed 
$k_1 = 0.4\,{\rm Mpc}^{-1}$. Different parameters, including $b_1$, $b_2$, 
$\sigma_p$, $n_g$ and volume $V$, at each redshift are taken following the choices in Table~1 of \citet{Yankelevich:2018uaz}. The matter power spectra with and without neutrinos at 
each redshift are obtained from \texttt{CLASS}.

As seen in the figure, the SNR in the $B_0^0$ multipole increases steadily with redshift.
The $B_2^0$ and $B_4^0$ multipoles show a different behavior: their SNR initially increases slightly, peaks around $z\sim1.0$, and then decreases at higher redshifts.
In contrast, the $B_2^1$ and $B_2^2$ multipoles show a mild but consistent increase of SNR with redshift, even though their overall values remain smaller compared to $B_0^0$ and $B_2^0$.
The $B_4^1$ multipole also displays a peak at intermediate redshift and a decline at higher $z$, similar to $B_4^0$.
Overall, the behavior of SNR with redshift varies significantly across different multipoles. The $B_0^0$ multipole becomes progressively stronger at higher redshifts, making it a robust probe, while for higher-order multipoles such as $B_2^0$ and $B_4^0$, the SNR peaks around $z\sim1$ and then declines. Interestingly, the $B_2^1$ and $B_2^2$ multipoles show a steady, albeit slower, growth with redshift, suggesting that certain higher-order terms might become relatively more important at later times.

\section{ Summary and Discussion}
\label{sec:discussions}

In this work, we have developed a detailed theoretical framework to model the impact of 
massive neutrinos on the redshift-space galaxy bispectrum, using a spherical harmonic 
multipole expansion. Unlike real-space analyses, where neutrino effects in the bispectrum 
remain at the sub-percent level, we find that redshift-space distortions amplify the neutrino 
signatures, making them potentially detectable with large-volume 
galaxy surveys such as \textit{Euclid}.

Our results indicate that different multipoles are affected to varying degrees. The multipoles $Q_0^0,Q_2^0,Q_4^0$ show the strongest sensitivity to neutrinos, particularly in linear triangle configurations near the squeezed limit. The maximum effect reaches up to  $\sim 1.4\%$ for lower multipoles  ($Q_L^m \leq Q_2^2 $) and exceeds $2\%$ in some higher multipoles ($Q_L^m \geq Q_4^0$) for a total neutrino mass of $\sum m_\nu =0.12 \, {\rm eV}$. These estimates account for changes in anisotropy due to neutrino-induced modifications of the growth rate and PT kernels under the Einstein-de Sitter (EdS) approximation. Odd multipoles also display peak deviations near linear/squeezed configurations, with $L=8$ showing notable patterns—although their amplitude is significantly smaller than that of the monopole. We also find that these effects are mildly sensitive to the non-linear bias parameter $\gamma_2$. For instance, in the $Q_0^0$ multipole, the peak residual shifts from equilateral to linear configurations as $\gamma_2$ increases from $-0.9$. Additionally, we consider the impact of galaxy peculiar velocities at small scales, known as the Finger-of-God (FoG) effect. Since FoG suppresses clustering, it leads to a systematic suppression in all multipoles.

We find that the largest neutrino-induced signatures arise in the squeezed and stretched triangle configurations. 
For detecting neutrino signatures, we make use of the signal-to-noise ratio (SNR) 
which is defined as the difference between bispectrum multipoles in the presence and absence of massive
neutrinos, divided by the expected variance of the multipoles with neutrinos. 
The $B_0^0$ (monopole) multipole shows the highest SNR, followed closely by the $B_2^0$ multipole. 
SNR forecasts suggest that these multipoles can achieve $>5\sigma$ detection significance 
in a \textit{Euclid} survey, even when realistic observational effects such as Finger-of-God (FoG) 
damping and shot noise are included. Higher-order multipoles, including $B_2^1$, $B_2^2$, $B_4^0$, and $B_4^1$, while having lower intrinsic SNRs, are still predicted to be detectable at a few$-\sigma$ level, especially when appropriate binning strategies are employed to boost cumulative signal strength.

We further explore the scale and redshift dependence of the SNR. The SNR grows systematically with increasing wavenumber $k_1$, which is defined to be the magnitude of the largest side of the bispectrum triangle in Fourier space, reflecting the larger number of available modes and the increasing relative importance of neutrino-induced suppression at smaller scales. For squeezed triangle configurations, we find that $B_4^0$ and $B_4^1$ multipoles, despite being suppressed at large scales, can reach remarkably high SNRs (e.g., $\sim60$ for $B_4^0$) at $k_1 \sim 0.8\,{\rm Mpc}^{-1}$.
In terms of redshift evolution, we find that the SNR in $B_0^0$ increases steadily with redshift, whereas the SNRs of $B_2^0$ and $B_4^0$ peak around $z \sim 1$ and decline thereafter. Interestingly, higher multipoles with $m=1$ and $m=2$ components ($B_2^1$ and $B_2^2$) exhibit a gradual increase with redshift, suggesting that they may play a more prominent role at higher redshifts.

The importance of this work lies in demonstrating that the redshift-space bispectrum, 
analyzed via its full multipole structure, provides a complementary and powerful avenue 
for probing the effects of massive neutrinos. While previous constraints on neutrino masses from large-scale structure have largely relied on the two-point correlation function or power spectrum, 
or on the bispectrum monopole, the higher bispectrum multipoles capture additional information, offering a route to enhance sensitivity to neutrino properties.

While the present analysis is based on tree-level perturbation theory, it sets a solid foundation for more refined modeling. Future work could incorporate one-loop corrections or effective field theory (EFT) treatments to extend the formalism to smaller scales where non-linearities become significant. Incorporating full survey effects — such as survey geometry, selection functions, fiber collisions, and imperfect knowledge of galaxy bias — will be necessary for truly robust forecasts. Furthermore, a joint analysis combining bispectrum multipoles with power spectrum multipoles could substantially improve constraints, leveraging the full statistical power of large galaxy surveys like \textit{Euclid}. 
We leave a comprehensive investigation of these aspects for future studies.

In summary, our results show that large galaxy surveys like \textit{Euclid} have the potential 
to detect and utilize higher-order statistics like the bispectrum to constrain fundamental physics, including the absolute neutrino mass scale. The bispectrum multipole expansion thus opens a new and complementary window for cosmological neutrino detection, advancing the frontier of large-scale structure cosmology.

\section*{Acknowledgements}
We sincerely thank Yvonne Y.Y. Wong for insightful discussions.
We gratefully acknowledge the use of the publicly available code \href{https://github.com/lesgourg/class_public}{\texttt{CLASS}}.  We also acknowledge the computational facilities of the Technology Innovation Hub, ISI Kolkata, and of the Pegasus cluster of the high-performance computing (HPC) facility at IUCAA, Pune. 
SP1 thanks CSIR for financial support through Senior Research Fellowship (File no. 09/093(0195)/2020-EMR-I).
DS acknowledges the support of the Canada $150$ Chairs program, the Fonds de recherche du Qu\'{e}bec Nature et Technologies (FRQNT) and Natural Sciences and Engineering Research Council of Canada (NSERC) joint NOVA grant, and the Trottier Space Institute Postdoctoral Fellowship program.
RS is supported by DST INSPIRE Faculty fellowship, India (Grant No.IFA19-PH231), NFSG and OPERA Research Grant from Birla Institute of Technology and Science, Pilani (Hyderabad Campus).
SP2 thanks  ANRF, Govt. of India for partial support through Project No.
CRG/2023/003984.

\section*{DATA AVAILABILITY}
\label{ref:data_av}
Numerical analyses presented in this work were primarily carried out using \texttt{Mathematica} and the publicly available \texttt{CLASS} code. The Mathematica notebooks used in these computations are available from the authors upon reasonable request in \href{https://github.com/sourav1729/RSD_bispectrum_with_neutrino}{https://github.com/sourav1729/RSD_bispectrum_with_neutrino}. The analytical expressions of all the multipole moments in presence of neutrinos are too lengthy to be reported in the main article. They can rather be found in \\ \href{https://github.com/sourav1729/supplementary_material}{https://github.com/sourav1729/supplementary_material}. 

\begin{figure*}
    \centering
    \subfloat[]{\includegraphics[width=0.55\columnwidth]{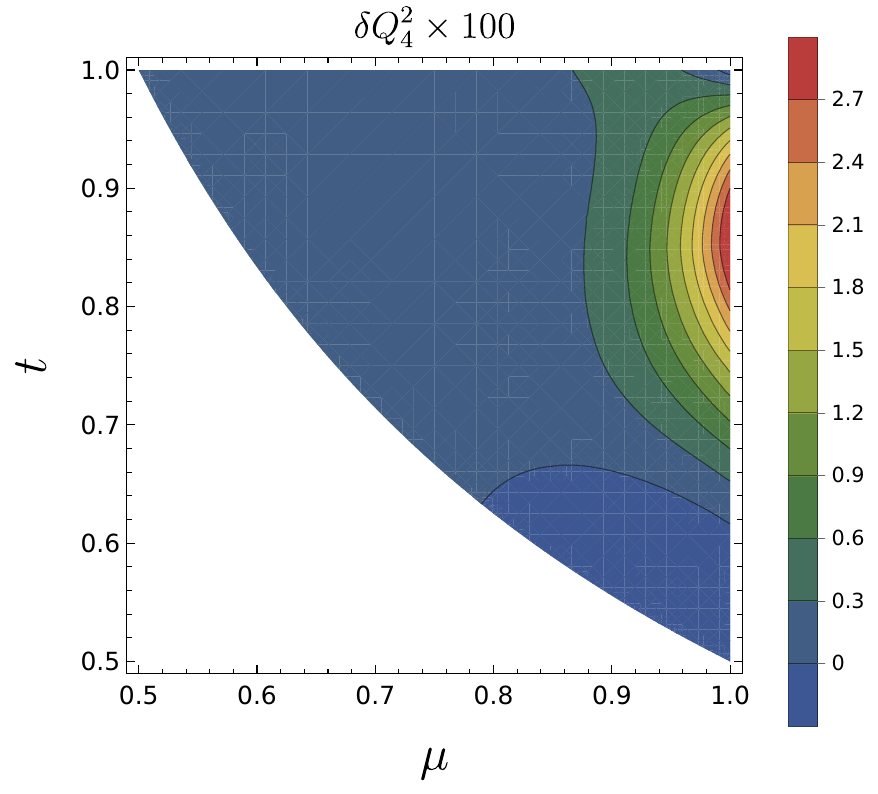}}
    \subfloat[]{\includegraphics[width=0.55\columnwidth]{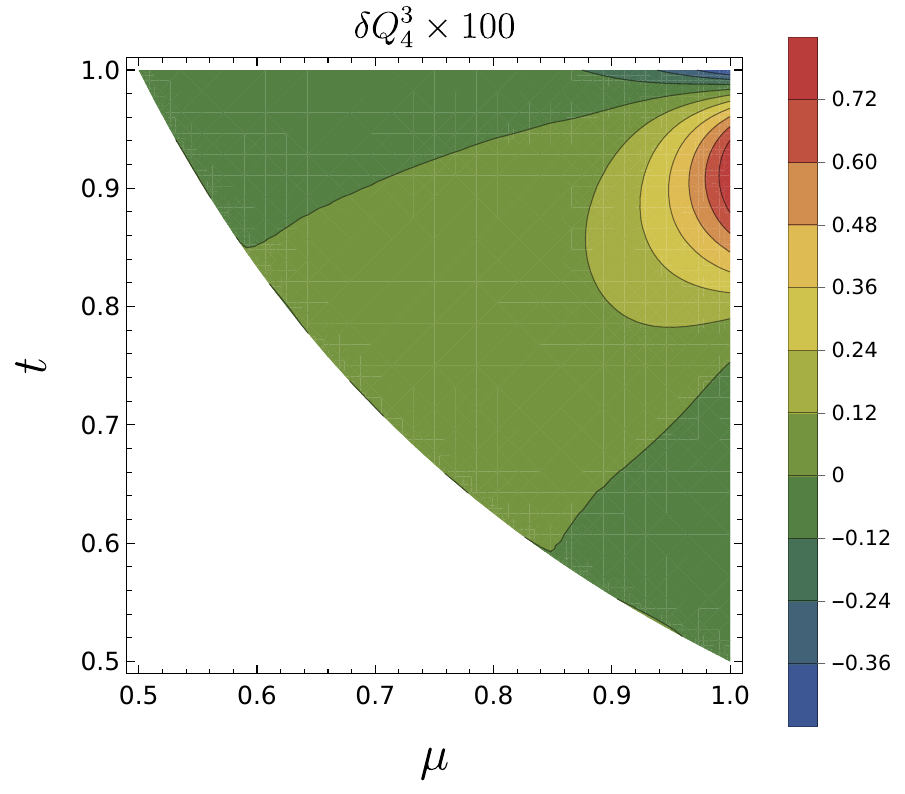}}
    \subfloat[]{\includegraphics[width=0.55\columnwidth]{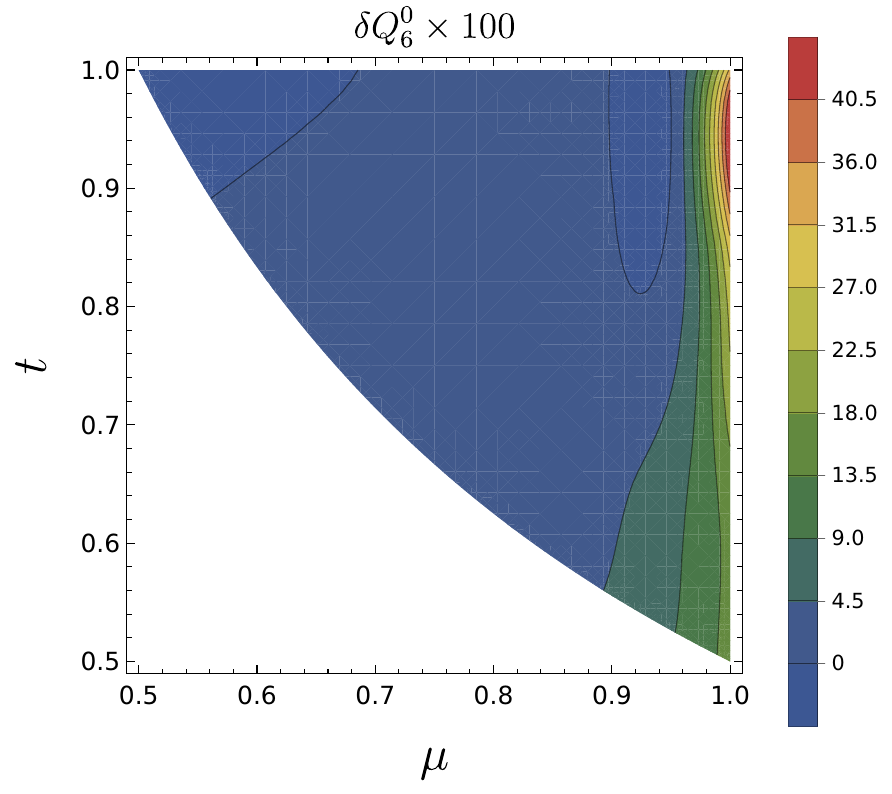}}
    \quad
    \subfloat[]{\includegraphics[width=0.55\columnwidth]{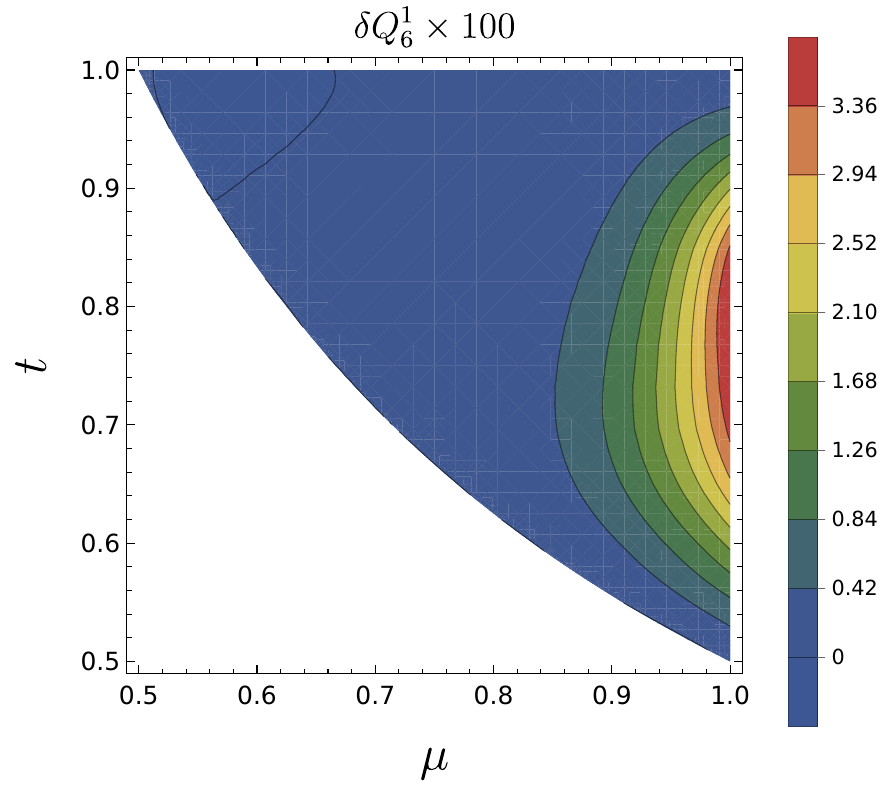}}
    \subfloat[]{\includegraphics[width=0.55\columnwidth]{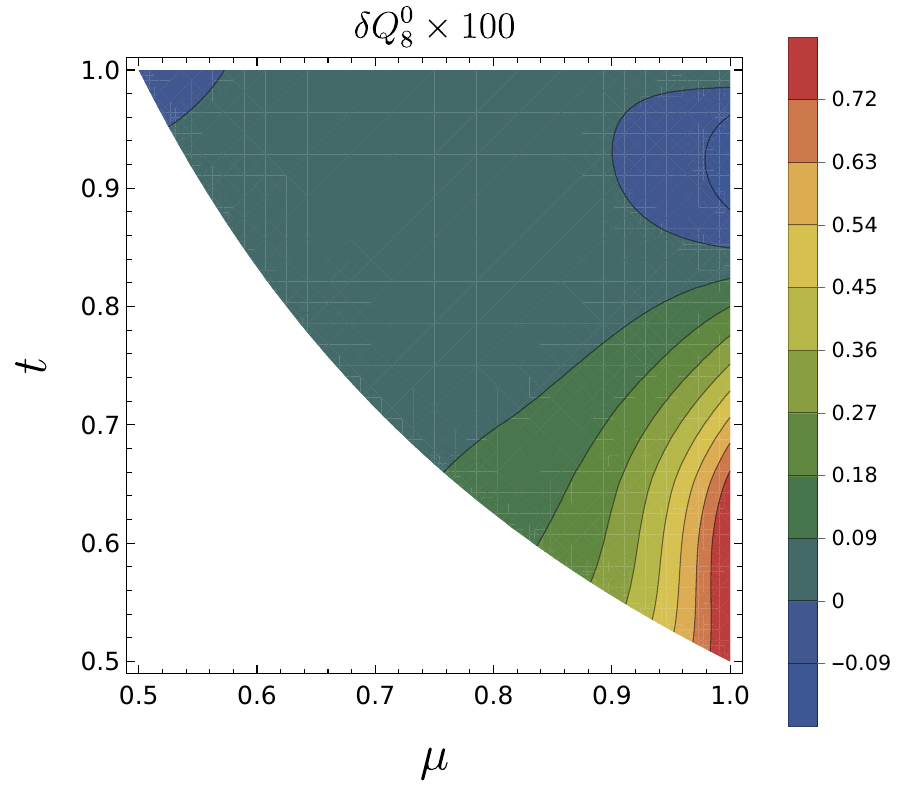}}
    \subfloat[]{\includegraphics[width=0.55\columnwidth]{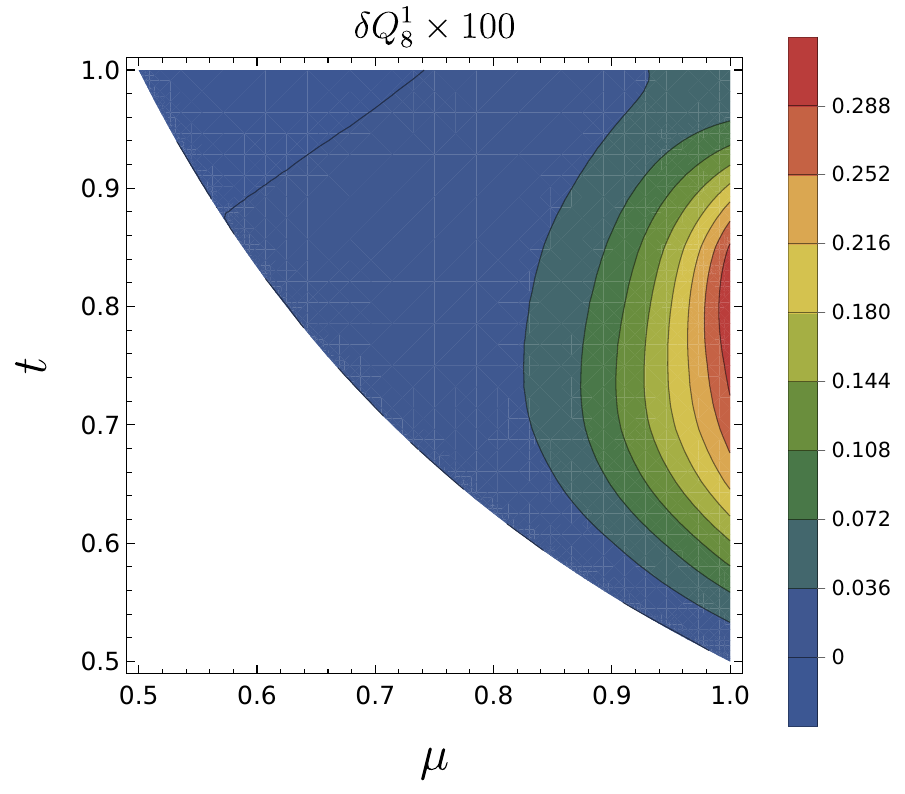}}
    \caption{Illustration of difference in the multipoles ($\delta Q_L^m$) in presence of neutrinos with $\sum m_\nu=0.12\, {\rm eV}$ in ($\mu-t$) plane. Each figure shows the corresponding multipole in its frame. We fix $k_1=0.2 \, {\rm Mpc}^{-1}$, $b_1=1.18$ and $\gamma_2=-0.9$ to generate all the figures. The matter power spectrum is evaluated at redshift $z=0.7$. From the plots it is evident that, maximum differences for $B_4^2$, $B_4^3$ and $B_6^0$ are located near squeezed triangle configuration. In $B_6^1$, $B_8^0$ and $B_8^1$ multipoles, the peaks are located near linear/stretched triangle shapes.}
    \label{fig:qlm_2}
\end{figure*}

\bibliographystyle{mnras}
\bibliography{biblio}

\begin{figure*}
    \centering
    \subfloat[]{\includegraphics[width=0.7\columnwidth]{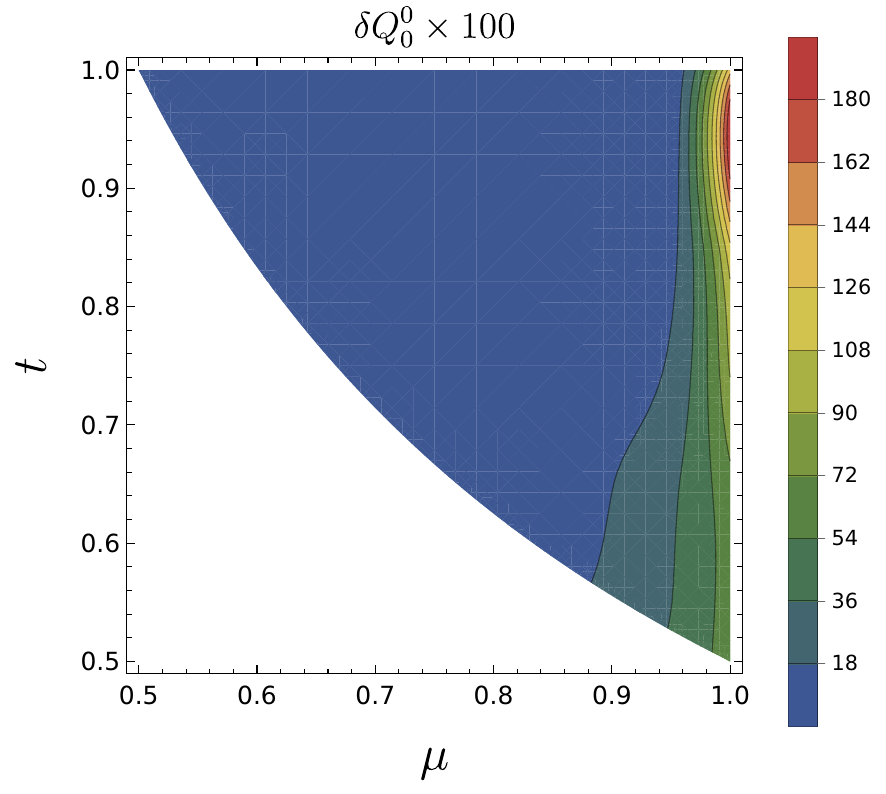}}
    \subfloat[]{\includegraphics[width=0.7\columnwidth]{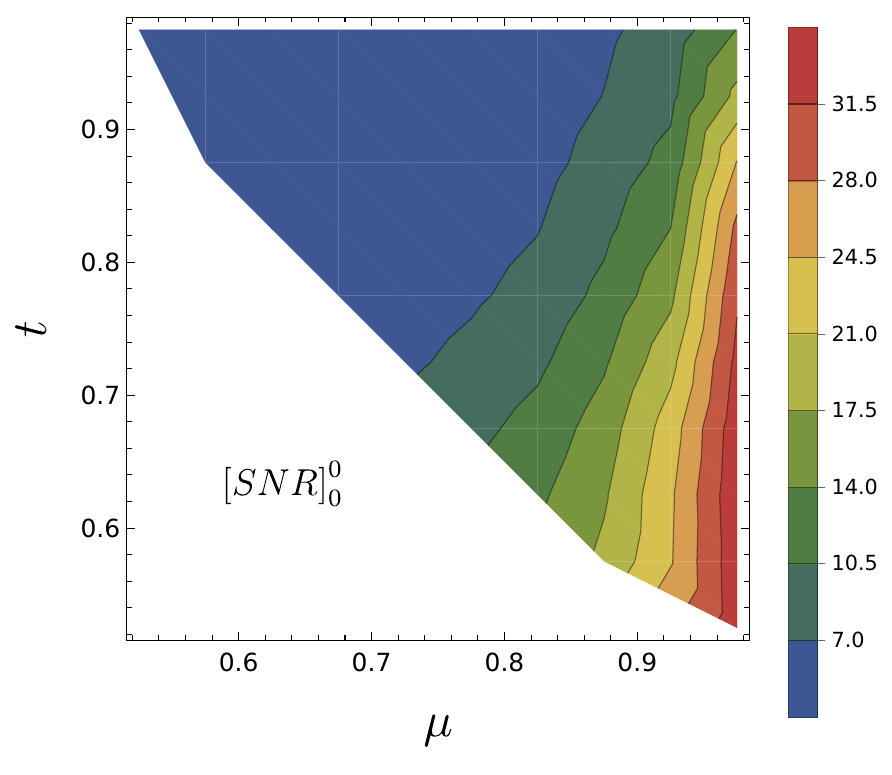}}
    \quad
    \subfloat[]{\includegraphics[width=0.7\columnwidth]{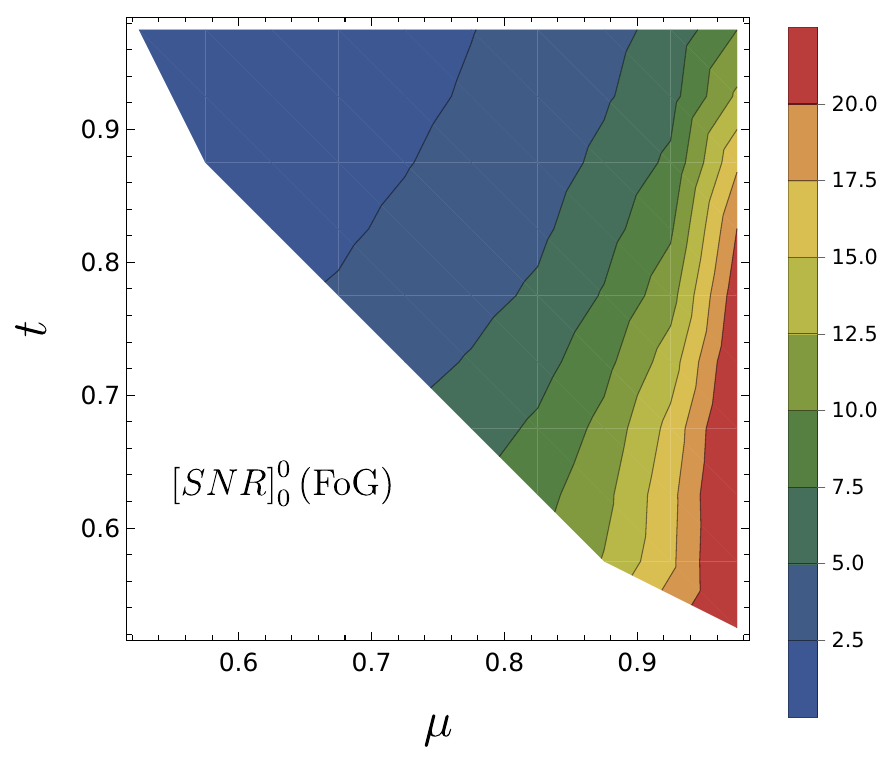}}
    \subfloat[]{\includegraphics[width=0.7\columnwidth]{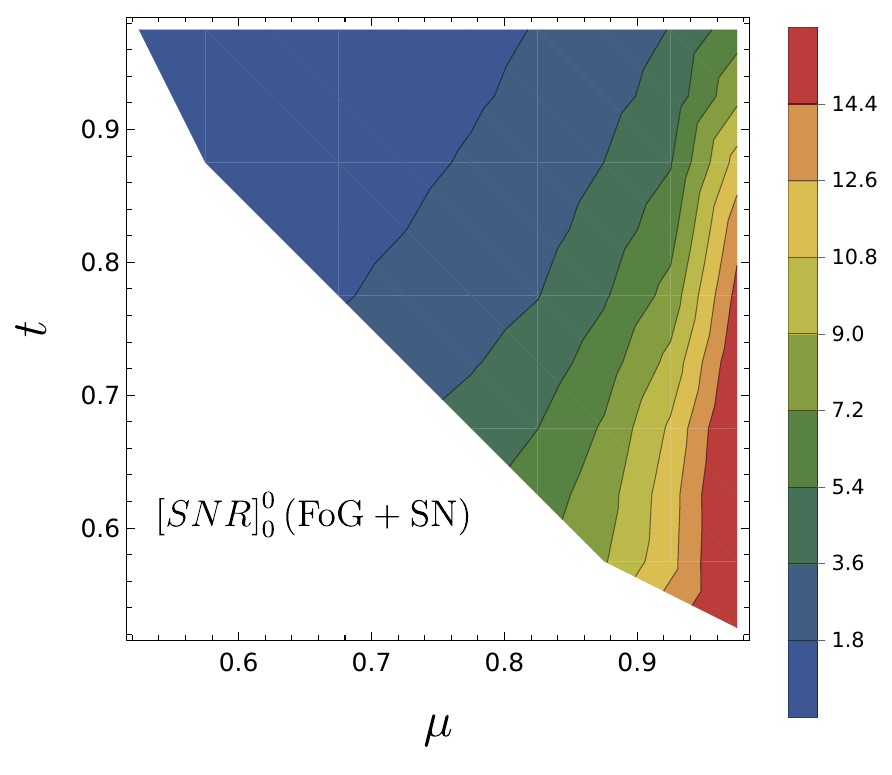}}
    \caption{Illustration of the effects of tidal bias on the monopole. In the \textit{upper left panel}, we show the difference in the $Q^0_0$ multipole when including the tidal bias term with $b_t = -0.4$. The \textit{upper right panel} displays the corresponding SNR maps for the $B^0_0$ multipole, computed without Fingers-of-God (FoG) or shot noise effects. In the \textit{lower panel}, we present the SNR maps in the $\mu$–$t$ space, now incorporating both FoG and shot noise contributions.}
    \label{fig:tidal_bias}
\end{figure*}

\appendix
\section{Effects on Higher multipoles}
\label{sec:appendixA}

In Fig.~\ref{fig:qlm_2}, the difference in the higher order multipoles due to the presence of neutrinos are presented. The results are shown for fixed $k_1=0.2\, {\rm Mpc}^{-1},\, b_1=1.18, \, \gamma_2 =-0.9$ at redshift $z=0.7$ for $\sum m_\nu=0.12 \, {\rm eV}$. We can see that, in $B_4^2$, $B_4^3$ and $B_6^0$ multipoles, maximum differences are located near the squeezed triangle configuration. In $B_6^1$, $B_8^0$ and $B_8^1$ multipoles, the maximum differences are located near the stretched triangle configuration. Since the absolute value of these multipoles in absence of neutrinos are themselves very small, we refrain from detailed study of these multipoles.

\section{Effects of Tidal bias}
\label{sec:appendixB}
In our analysis, we retain bias contributions up to the second-order non-linear bias term $b_2$, while neglecting the tidal bias $b_t$, although both appear at the same perturbative order in the galaxy bias expansion (Eq.~\eqref{eq:galaxy_bias}). The inclusion of tidal bias modifies the bispectrum by introducing additional triangle-dependent contributions beyond the standard $\tilde{F}_2$ and $\tilde{G}_2$ kernels. For completeness, we present selected results incorporating the tidal bias.

We focus on the monopole $B^0_0$, which exhibits the highest signal-to-noise ratio (SNR), although the analysis can be extended to higher multipoles. The results are summarized in Fig.\ref{fig:tidal_bias}, using a fiducial value of $b_t = -0.4$, the largest among the values quoted in Ref.\citep{Yankelevich:2018uaz}. Even for this relatively high value, the tidal bias produces only a marginal change in $B^0_0$ compared to the case without tidal bias shown in Fig.~\ref{fig:qlm_1}. The impact on higher multipoles is found to be even smaller.

The SNR for $B^0_0$ is also slightly reduced when tidal bias is included, but the effect remains significantly small. Given that $b_t$ can be approximated as $b_t \approx -\frac{2}{7}(b_1 - 1)$, its magnitude is expected to be small for typical low-redshift galaxy populations where $b_1 \sim 1$. This supports the robustness of our results even without the inclusion of the tidal bias term in the main analysis.

\end{document}